\documentclass[twocolumn]{aastex7}

\usepackage{amsmath}

\shorttitle{JADES Clumps}
\shortauthors{Zhu et al.}

\begin{document}


\title{Clumps in High-Redshift Galaxies: Mass Scaling and Radial Trends from JADES}

\author[0000-0003-3307-7525]{Yongda Zhu} \thanks{JASPER Scholar}
\affiliation{Steward Observatory, University of Arizona, 933 North Cherry Avenue, Tucson, AZ 85721, USA}
\email[show]{yongdaz@arizona.edu}

\author[0000-0002-7893-6170]{Marcia J. Rieke}
\affiliation{Steward Observatory, University of Arizona, 933 North Cherry Avenue, Tucson, AZ 85721, USA}
\email{mrieke@arizona.edu}

\author[0000-0001-7673-2257]{Zhiyuan Ji} \thanks{JASPER Scholar}
\affiliation{Steward Observatory, University of Arizona, 933 North Cherry Avenue, Tucson, AZ 85721, USA}
\email{}

\author[0000-0002-8651-9879]{Andrew J.\ Bunker}
\affiliation{Department of Physics, University of Oxford, Denys Wilkinson Building, Keble Road, Oxford OX1 3RH, UK}
\email{}

\author[0000-0001-6301-3667]{Courtney Carreira}
\affiliation{Department of Astronomy and Astrophysics, University of California, Santa Cruz, 1156 High Street, Santa Cruz, CA 95064, USA}
\email{ccarreir@ucsc.edu}

\author[0000-0002-9708-9958]{A. Lola Danhaive}
\affiliation{Kavli Institute for Cosmology, University of Cambridge, Madingley Road, Cambridge, CB3 0HA, UK}
\affiliation{Cavendish Laboratory, University of Cambridge, 19 JJ Thomson Avenue, Cambridge, CB3 0HE, UK}
\email{ald66@cam.ac.uk}

\author[0009-0009-8105-4564]{Qiao Duan}
\affiliation{Kavli Institute for Cosmology, University of Cambridge, Madingley Road, Cambridge, CB3 0HA, UK}
\affiliation{Cavendish Laboratory, University of Cambridge, 19 JJ Thomson Avenue, Cambridge, CB3 0HE, UK}
\email{qd231@cam.ac.uk}

\author[0000-0003-1344-9475]{Eiichi Egami}
\affiliation{Steward Observatory, University of Arizona, 933 North Cherry Avenue, Tucson, AZ 85721, USA}
\email{}

\author[0000-0002-2929-3121]{Daniel J.\ Eisenstein}
\affiliation{Center for Astrophysics $|$ Harvard \& Smithsonian, 60 Garden St., Cambridge MA 02138 USA}
\email{}

\author[0000-0003-4565-8239]{Kevin Hainline}
\affiliation{Steward Observatory, University of Arizona, 933 North Cherry Avenue, Tucson, AZ 85721, USA}
\email{kevinhainline@arizona.edu}

\author[0000-0002-9280-7594]{Benjamin D.\ Johnson}
\affiliation{Center for Astrophysics $|$ Harvard \& Smithsonian, 60 Garden St., Cambridge MA 02138 USA}
\email{}

\author[0009-0003-5402-4809]{Zheng Ma}
\affiliation{Steward Observatory, University of Arizona, 933 North Cherry Avenue, Tucson, AZ 85721, USA}
\email{}

\author[0000-0001-8630-2031]{Dávid Puskás}
\affiliation{Kavli Institute for Cosmology, University of Cambridge, Madingley Road, Cambridge, CB3 0HA, UK}
\affiliation{Cavendish Laboratory, University of Cambridge, 19 JJ Thomson Avenue, Cambridge, CB3 0HE, UK}
\email{dp670@cam.ac.uk}

\author[0000-0003-2303-6519]{George H.\ Rieke}
\affiliation{Steward Observatory, University of Arizona, 933 North Cherry Avenue, Tucson, AZ 85721, USA}
\email{grieke@arizona.edu}

\author[0000-0002-5104-8245]{Pierluigi Rinaldi}
\affiliation{AURA for the European Space Agency (ESA), Space Telescope Science Institute, 3700 San Martin Dr., Baltimore, MD 21218, USA}
\email{}

\author[0000-0002-4271-0364]{Brant Robertson}
\affiliation{Department of Astronomy and Astrophysics, University of California, Santa Cruz, 1156 High Street, Santa Cruz, CA 95064, USA}
\email{brant@ucsc.edu}

\author[0000-0002-8224-4505]{Sandro Tacchella}
\affiliation{Kavli Institute for Cosmology, University of Cambridge, Madingley Road, Cambridge, CB3 0HA, UK}
\affiliation{Cavendish Laboratory, University of Cambridge, 19 JJ Thomson Avenue, Cambridge, CB3 0HE, UK}
\email{st578@cam.ac.uk}

\author[0000-0003-4891-0794]{Hannah \"Ubler}
\affiliation{Max-Planck-Institut f\"ur extraterrestrische Physik (MPE), Gie{\ss}enbachstra{\ss}e 1, 85748 Garching, Germany}
\email{hannah@mpe.mpg.de}

\author[0000-0001-6917-4656]{Natalia C. Villanueva}
\affiliation{Department of Astronomy, The University of Texas at Austin, Austin, TX, USA}
\email{nataliavillanueva@utexas.edu}

\author[0000-0003-2919-7495]{Christina C.\ Williams}
\affiliation{NSF National Optical-Infrared Astronomy Research Laboratory, 950 North Cherry Avenue, Tucson, AZ 85719, USA}
\email{christina.williams@noirlab.edu}

\author[0000-0001-9262-9997]{Christopher N.\ A.\ Willmer}
\affiliation{Steward Observatory, University of Arizona, 933 North Cherry Avenue, Tucson, AZ 85721, USA}
\email{cnaw@arizona.edu}

\author[0000-0002-8876-5248]{Zihao Wu}
\affiliation{Center for Astrophysics $|$ Harvard \& Smithsonian, 60 Garden St., Cambridge MA 02138 USA}
\email{zihao.wu@cfa.harvard.edu}

\author[0000-0002-1574-2045]{Junyu Zhang}
\affiliation{Steward Observatory, University of Arizona, 933 North Cherry Avenue, Tucson, AZ 85721, USA}
\email{}

\begin{abstract}
Massive star-forming clumps are a prominent feature of high-redshift galaxies and are thought to trace gravitational fragmentation, feedback, and bulge growth in gas-rich disks. We present a statistical analysis of \added{clumps} in $\sim$3600 galaxies spanning $2 \lesssim z \lesssim 8$ from deep JWST/NIRCam imaging in the JADES GOODS--South field. Clumps are identified as residual features after subtracting smooth S\'ersic profiles, enabling a uniform, rest-frame optical census of sub-galactic structure. We characterize their physical properties, size--mass relations, and spatial distributions to constrain models of sub-galactic structure formation and evolution. We find that clumps in our sample are typically low-mass ($10^{\sim7-8}M_\odot$), actively star-forming, and show diverse gas-phase metallicity, dust attenuation, and stellar population properties. Their sizes and average pairwise separations increase with cosmic time (toward lower redshift), consistent with inside-out disk growth. The clump mass function follows a power law with slope $\alpha = -1.50_{-0.17}^{+0.19}$, consistent with fragmentation in turbulent disks. We find a deficit of relatively young clumps near galaxy centers and a radial transition in the size--mass relation: outer clumps exhibit steeper, near-virial slopes ($R_{\rm e}\propto M_*^{\sim 0.3}$), while inner clumps follow flatter trends ($R_{\rm e}\propto M_*^{\sim 0.2}$), consistent with structural evolution via migration or disruption. These results provide new constraints on the formation, survival, and dynamical evolution of clumps, highlighting their role in shaping galaxy morphology during the peak of cosmic star formation.
\end{abstract}


\section{Introduction}

High-redshift galaxies often exhibit irregular, clumpy morphologies, characterized by compact substructures that deviate from smooth light profiles. These \added{clumps} are widely interpreted as sites of intense star formation embedded in gas-rich, turbulent disks \citep[e.g.,][]{elmegreen_chain_2004, elmegreen_clumpy_2009,romeo_toomre-like_2010,romeo_larsons_2014, forster_schreiber_constraints_2011, guo_multi-wavelength_2012, wuyts_smoother_2012,ikeda_formation_2025}. Clumps are particularly common during the epoch of cosmic noon ($z \sim 1$-3), when galaxies are rapidly assembling stellar mass and the interstellar medium (ISM) is dynamically unstable \citep{forster_schreiber_constraints_2011, genzel_sins_2011, tacchella_evidence_2015, shibuya_morphologies_2016}. Recent JWST observations confirm the ubiquity of clumpy substructures in the rest-frame near-infrared, revealing star-forming clumps that were unresolved or missed in previous optical studies \citep[e.g.,][]{kalita_near-ir_2025, kalita_clumps_2025, vega_fraction_2025, sok_stellar_2025, mawatari_rioja_2025}.

Several formation mechanisms have been proposed. A leading model invokes violent disk instability (VDI; e.g., \citealp{krumholz_unified_2018}), in which cold gas accretion and high gas fractions drive fragmentation in marginally stable disks \citep{dekel_cold_2009, ceverino_high-redshift_2010, romeo_toomre-like_2010, romeo_larsons_2014, bournaud_long_2014, inoue_non-linear_2016, mandelker_giant_2017, oklopcic_giant_2017}. Alternative pathways include minor or major mergers, satellite accretion, and clumps triggered by tidal interactions or shocks \citep[e.g.,][]{immeli_gas_2004, puech_forming_2009, calabro_merger_2019, nakazato_merger-driven_2024}. Some clumps may be accreted satellites or merging dwarf companions, consistent with the elevated merger rates at these redshifts \citep[e.g.,][]{duan_galaxy_2025, puskas_constraining_2025}. Discriminating between these formation channels remains a key challenge in understanding the origin of internal substructures.

Clumps may also play a dynamic role in shaping the internal conditions of high-redshift galaxies. In gas-rich disks, gravitational instabilities and rapid inflow can trigger the formation of massive clumps whose feedback and gravitational potential help sustain elevated turbulence in the surrounding medium. Clump-driven stirring has been linked to high ionized-gas velocity dispersions in both simulations and observations \citep[e.g.,][and references therein]{ubler_evolution_2019}, and recent measurements show that dispersions continue to rise toward high redshift, where gas fractions and instability growth rates are highest \citep{danhaive_dawn_2025}. Galaxies with potentially strong feedback, as probed by extended line emission in JWST/NIRCam medium bands, frequently show irregular and clumpy morphologies \citep{zhu_systematic_2025}. Simulations further suggest that massive clumps may survive feedback and migrate inward through dynamical friction or gravitational torques, contributing to bulge growth and central mass buildup \citep[e.g.,][]{ceverino_rotational_2012, mandelker_population_2014}. Testing these scenarios observationally requires measuring how clump properties vary with galactocentric distance---particularly their stellar masses and structural scaling relations---as demonstrated in several lower-redshift ($z \lesssim 2.5$) studies \citep[e.g.,][]{guo_multi-wavelength_2012, tadaki_nature_2014, shibuya_morphologies_2016}.

Previous studies of clumps have largely relied on \textit{Hubble Space Telescope} (\textit{HST}) imaging at rest-frame UV wavelengths or on small samples of strongly lensed galaxies \citep[e.g.,][]{guo_clumpy_2015, livermore_resolved_2015, zanella_extremely_2015, zanella_contribution_2019}. While these data provided early constraints on clump sizes and star formation activity, they are limited by spatial resolution, sample size, and selection biases. JWST now enables high-resolution, rest-frame optical imaging for large, unlensed galaxy samples. For example, the JWST Advanced Deep Extragalactic Survey (JADES; \citealp{rieke_jades_2023, eisenstein_overview_2023, eisenstein_jades_2025}) delivers ultra-deep NIRCam \citep{rieke_performance_2023} imaging of thousands of galaxies across a wide redshift baseline, offering the first opportunity to statistically characterize clumps over $z = 2$--8.

Recent JWST-based studies have begun to quantify clump prevalence across cosmic time. Using a uniform definition of clumpiness based on residual asymmetries, \citet{claeyssens_star_2023, vega_fraction_2025} find that the clumpy galaxy fraction increases with time from $\sim$20\% at $z > 6.5$ to nearly 80\% at $z \sim 2.8$, consistent with an evolving population of unstable, gas-rich disks. In this work, we identify \added{clumps} in the JADES imaging as residual substructures after subtracting smooth S\'ersic models and isolate them via watershed segmentation. This residual-based method captures both bright star-forming knots and lower-contrast irregularities, enabling consistent clump detection across a broad dynamic range. We analyze $\sim$3600 galaxies in the GOODS--South field to study the demographics, physical properties, spatial distribution, and evolution of clumps across cosmic time.

We aim to address several open questions that remain central to understanding clump formation and evolution at $z > 2.5$. How does clump frequency vary with redshift and host morphology? Do clumps follow coherent size--mass relations, and are these trends dependent on galactocentric distance? Are there radial gradients in age or structure that point to clump migration? And what does the clump mass function reveal about the physics of disk fragmentation and feedback?

This paper is organized as follows. Section~\ref{sec:data} describes the imaging data, sample selection, and S\'ersic modeling. In Section~\ref{sec:method}, we outline the residual-based clump detection pipeline. Sections~\ref{sec:clumpy_demo}--\ref{sec:mass_function} present results on clump frequency, physical properties, size--mass scaling, and mass functions. Section~\ref{sec:radial} examines radial trends and possible migration signatures. We discuss the implications in Section~\ref{sec:discussion} and summarize our conclusions in Section~\ref{sec:summary}. Appendix~\ref{sec:app_completeness} presents clump detection completeness tests, and Appendix~\ref{sec:app_gini_test} explores the role of galaxy morphology in driving clump multiplicity. Throughout this paper, we use a flat $\Lambda$CDM cosmology with $\Omega_{\rm m} = 0.3$, $\Omega_\Lambda = 0.7$, and $H_0 = 70$ km s$^{-1}$ Mpc$^{-1}$.

\section{Data and Method} \label{sec:data}

\added{In this section, we describe the data and methods used to identify clumps. We note that ``clumps'' in this paper refer to any clumpy substructure standing above the smooth S\'ersic galaxy profile. We do not attempt to remove features that may be (partly) associated with bulges or spiral arms. Visual inspection indicates that only $\sim 6\%$ of clumps exhibit elongated or extended morphologies that could plausibly be linked to these components, and they do not significantly affect our main results.}

\begin{figure*}
\centering
\includegraphics[width=\textwidth]{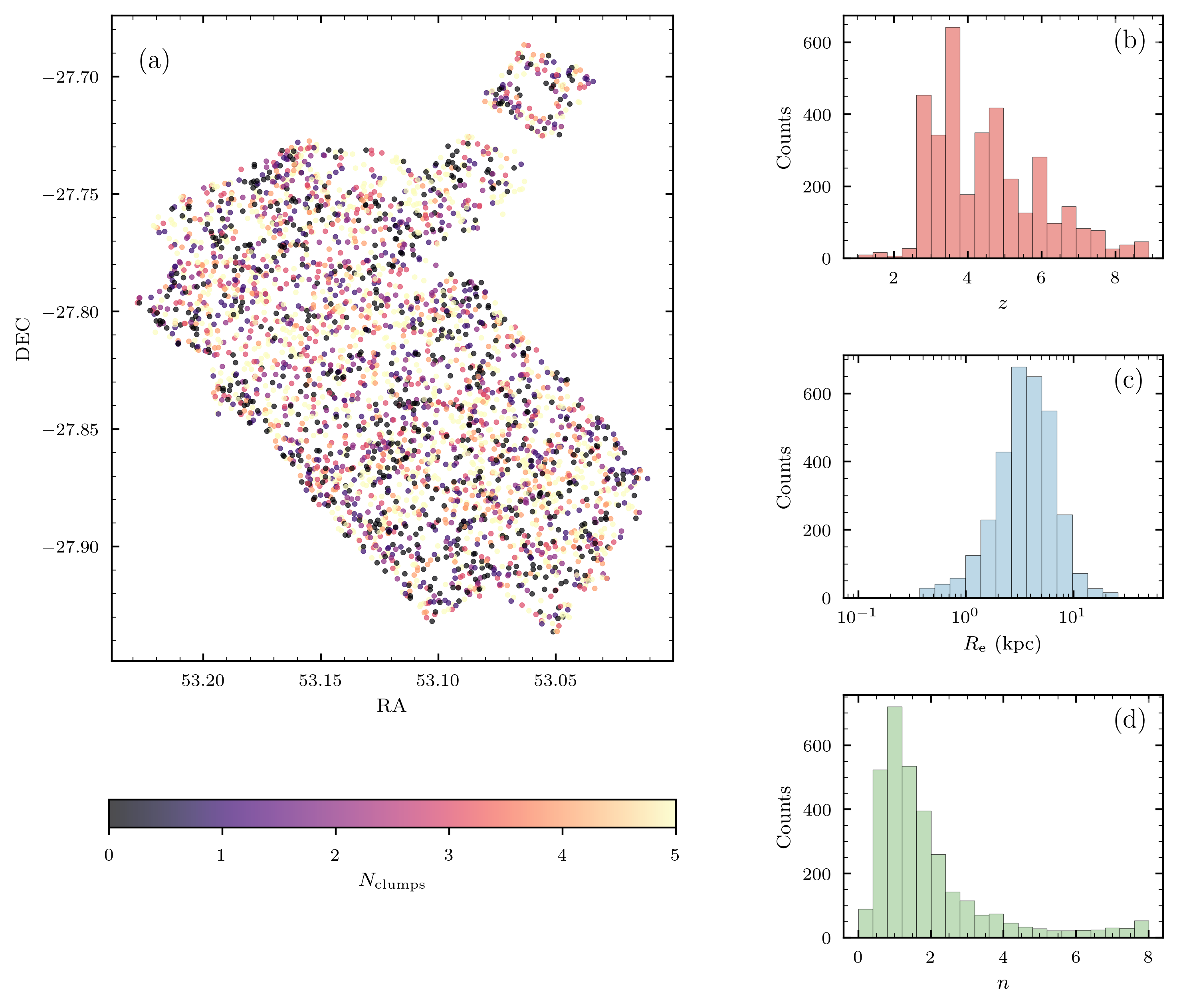}
\caption{
Overview of the parent galaxy sample in the JADES GOODS--South field. 
(a) Sky distribution of galaxies color-coded by the number of identified clumps ($N_{\mathrm{clumps}}$), with clumpy galaxies highlighted. 
(b) Redshift distribution of the full sample. 
(c) Distribution of effective radii ($R_{\mathrm{e}}$) in kpc.
(d) Distribution of S\'ersic indices ($n$) for the full sample. 
}
\label{fig:skymap_sample}
\end{figure*}

\subsection{Sample and Imaging}
We use deep near-infrared imaging from the JADES DR2 GOODS--South field \citep{eisenstein_jades_2025}. The dataset combines ultra-deep JWST/NIRCam observations with extensive spectroscopic and photometric legacy coverage \citep{rieke_jades_2023,eisenstein_overview_2023,eisenstein_jades_2025,hainline_cosmos_2024,bunker_jades_2024,deugenio_jades_2025,curtis-lake_jades_2025,scholtz_jades_2025}. Our final sample includes 3586 galaxies spanning $z = 1.0$--8.9 ($z_{95\%} = 2.6$--8.2), selected from a parent sample of 4000 randomly chosen galaxies after excluding objects with unreliable photometry or poor structural fits (see Section~\ref{sec:sersic}). Stellar masses range from $\log(M_\star / M_\odot) \simeq 7$-10.5. We use six NIRCam broadband filters, F115W, F150W, F200W, F277W, F356W, and F444W, covering rest-frame optical wavelengths out to $z \sim 8$, for structural analysis due to their high signal-to-noise ratios. All available JADES NIRCam imaging as well as NIRCam medium-band imaging from JEMS \citep{williams_jems_2023} and FRESCO \citep[][]{oesch_jwst_2023} are used for spectral energy distribution (SED) fitting (see Section \ref{sec:sed_properties}). Redshifts for all galaxies in our sample are taken from high-quality photometric estimates. We adopt the JADES photo-$z$ catalog \citep{hainline_cosmos_2024}, which is derived using the \texttt{EAZY} code \citep{brammer_eazy_2008}. JADES photometric redshifts closely follow NIRSpec and NIRCam grism spectroscopic redshifts ($z_{\mathrm{spec}}$), with typical scatter of $\Delta z / (1+z_{\mathrm{spec}}) \lesssim 0.1$ and minimal bias \citep[e.g.,][]{helton_identification_2024, zhu_smiles_2025}. These uncertainties have a negligible impact on the structural and clump-related trends explored in this work. An overview of the sample is shown in Figure \ref{fig:skymap_sample}.

\subsection{Structural Modeling and Residuals} \label{sec:sersic}

\begin{figure*}[!ht]
\centering
\includegraphics[width=1.0\textwidth]{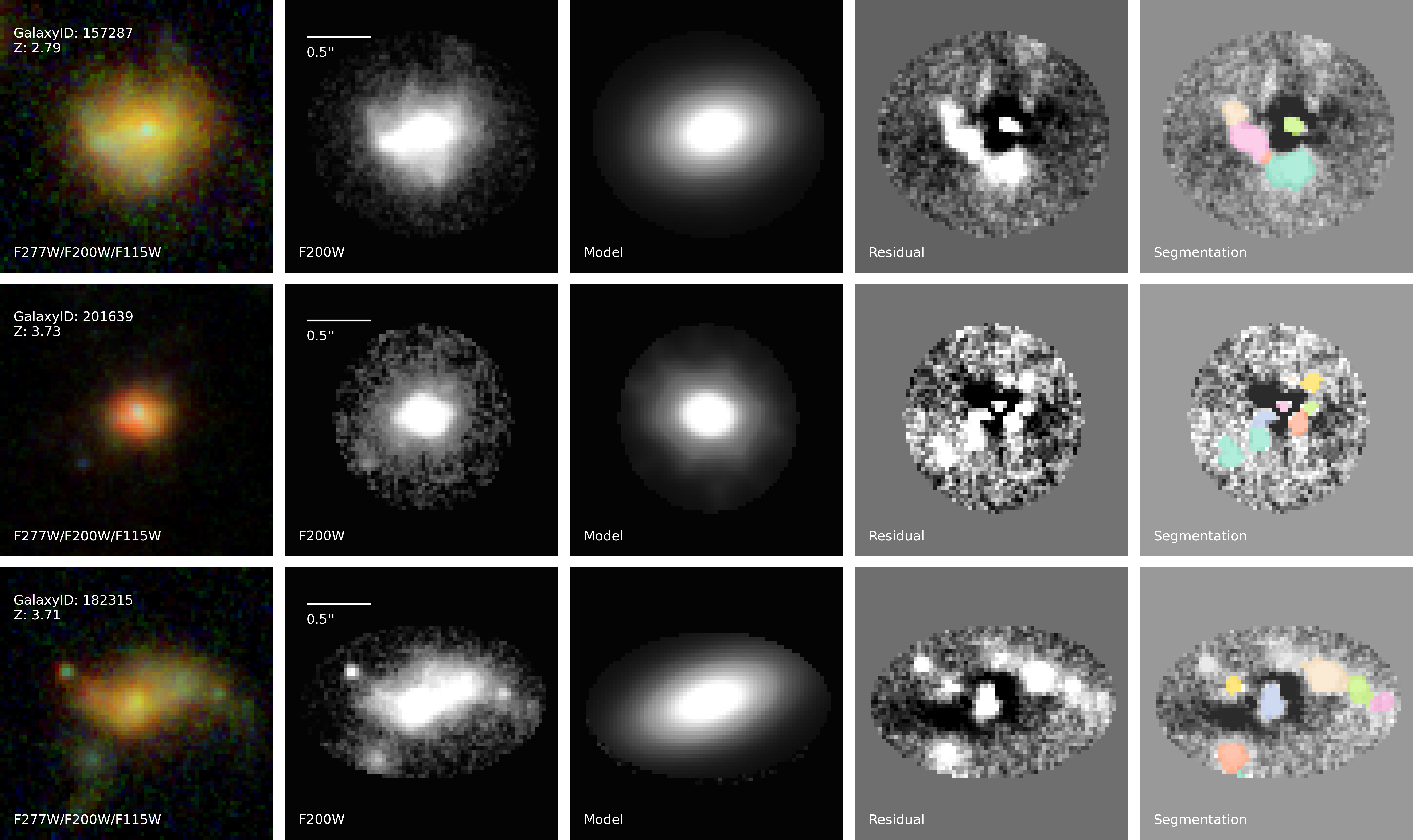}
\caption{
Clump identification procedure for three example galaxies. From left to right: (1) three-color NIRCam image using F277W (red), F200W (green), and F115W (blue); (2) F200W image used for S\'ersic profile fitting and clump detection (shown within the JADES \textsc{kron} aperture); (3) best-fit PSF-convolved S\'ersic model generated using \texttt{PySersic}; (4) residual image after subtracting the model; (5) segmentation map from watershed clump detection, with clump regions shown in colors.
}
\label{fig:clump_method}
\end{figure*}

We model each galaxy with two-dimensional S\'ersic profiles using \texttt{PySersic} \citep{pasha_pysersic_2023}, fitting each filter independently. All fits are visually inspected, and we exclude those with catastrophic failures or masking artifacts. Structural parameters of the galaxies include effective radius ($R_{\rm e}$), S\'ersic index ($n$), and total flux. Residual images are created by subtracting the best-fit model from the original image, isolating non-smooth and \added{clumps} (Figure~\ref{fig:clump_method}).

\subsection{Clump Detection and Selection} \label{sec:method}

We identify clumps from residual images using {\tt watershed} segmentation implemented in {\tt scikit-image} \citep{van_der_walt_scikit-image_2014}. Clumps are detected in the NIRCam broadband filter closest to rest-frame 5000\,\AA\ at each galaxy's redshift. To suppress pixel-scale noise, residual images are first smoothed with a Gaussian filter with $\sigma = 1$ pixel ($\sim0.03\arcsec$). The noise level is estimated from the smoothed residual image, and a binary detection mask is defined by thresholding at $3\times$ the background RMS, measured across the local $3\arcsec \times 3\arcsec$ cutout region. This mask highlights significant residual features while suppressing low-level fluctuations. A 2D Euclidean distance transform is then applied to the binary mask, and local maxima are identified with a minimum separation of 5 pixels. These peaks serve as markers for the {\tt watershed} algorithm, which partitions the residual image into catchment basins around each peak. This method naturally separates adjacent clumps in crowded systems and provides a non-parametric, shape-independent definition of clump boundaries \citep[e.g.,][]{meyer_topographic_1994}.

Each segmented region is treated as a candidate clump. To exclude spurious detections, we require that the total flux in the residual image within the segmented area exceeds 20\% of the total flux in the original (pre-subtraction) image over the same region. This ensures that the detected clump contributes significantly beyond the smooth S\'ersic model. Varying this threshold between 10--30\% does not affect our main results. Additionally, we discard regions with fewer than five connected pixels to eliminate noise spikes and artifacts. 

For each clump, we measure the effective radius directly from the segmentation mask and extract photometry in the NIRCam band closest to rest-frame 5000\,\AA\ at the galaxy's redshift. The number of galaxies contributing size measurements in each band is 18 (F115W), 29 (F150W), 1449 (F200W), 1165 (F277W), 671 (F356W), and 254 (F444W). These measurements are confined to the clump region defined by the residual segmentation map and are not deblended from the host light. As shown by \citet{kalita_near-ir_2025}, clumps in high-redshift disks have rest-UV and UVJ colors similar to their hosts, so contamination from surrounding light is not expected to strongly bias the inferred stellar population properties.

We do not attempt to model the morphology or light profile of individual clumps. The residual-based approach identifies statistically significant flux excesses relative to a smooth S\'ersic model, enabling uniform detection across a large sample but making the results sensitive to residual structure that deviates from the global fit. In principle, over- or under-subtraction of the host profile can affect clumps at all radii; however, the effect is most pronounced near galaxy centers where the light profile is steep and small mismatches in the S\'ersic fit lead to large residuals. As a result, off-center clumps are generally recovered more robustly, whereas centrally embedded clumps can be partially subtracted along with the host profile. The smallest measurable sizes are limited by the PSF, typically $\sim$0\farcs05 in F150W and $\sim$0\farcs1 in F356W. As demonstrated by our completeness tests (Appendix~\ref{sec:app_completeness}), brighter and more compact clumps are more easily recovered, whereas faint or extended central clumps are detected less efficiently. Our goal in this work is not to construct a complete clump census, but to characterize the structural and stellar population properties of reliably detected \added{clumps}. A comprehensive assessment of the clumpy-galaxy fraction and in JADES their connection with mergers will be presented in D.~Puskas et al.\ (in preparation).

\section{Results} \label{sec:results}
\added{In this section, we present the statistical properties of clumps identified in the JADES sample, including their frequency, stellar population characteristics, size--mass scaling relations, mass function, and radial structural trends.}

\begin{figure*}[!ht]
\centering
\includegraphics[width=0.8\linewidth]{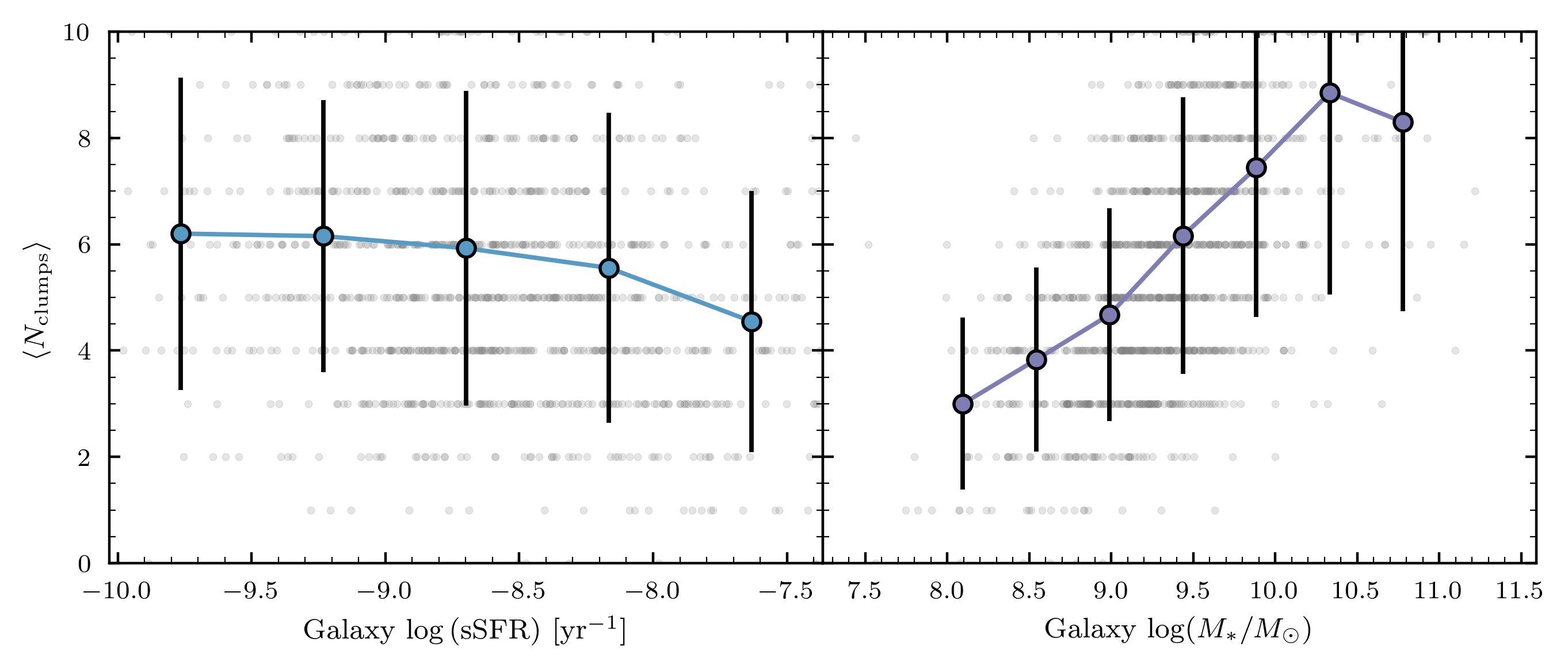}
\includegraphics[width=0.8\linewidth]{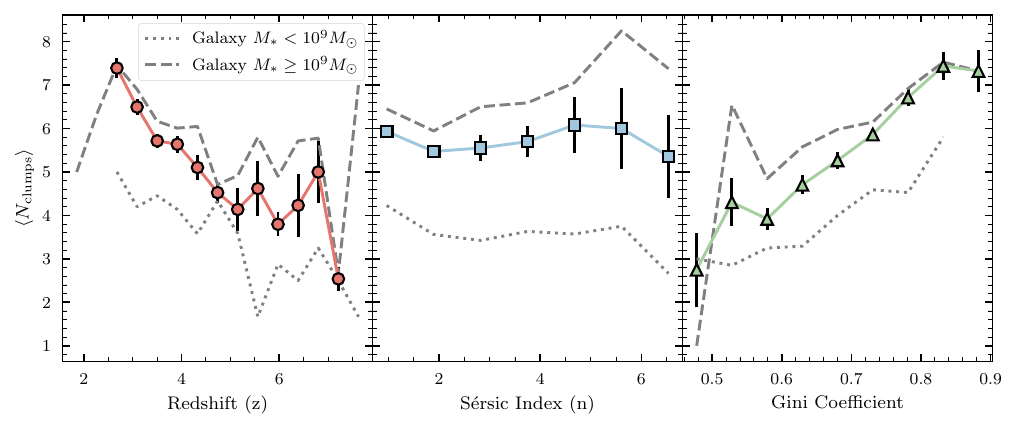}
\caption{
Average number of \added{clumps} per galaxy as a function of galaxy properties. 
Top row shows $\langle N_{\rm clumps} \rangle$ as a function of galaxy specific star formation rate (left) and stellar mass (right). Stellar mass exhibits a stronger correlation with clump number than specific star formation rate. Error bars in the top panels indicate the standard deviation of $N_{\rm clumps}$ within each bin. 
Bottom row shows $\langle N_{\rm clumps} \rangle$ as a function of redshift (left), S\'ersic index (middle), and Gini coefficient (right). Error bars in the bottom panels represent the 16th to 84th percentile range of the binned averages. To isolate the role of stellar mass, dotted and dashed lines indicate low mass ($M_* < 10^9\,M_\odot$) and high mass ($M_* \geq 10^9\,M_\odot$) galaxies, respectively. The overall trends with redshift and morphology remain consistent across stellar mass bins.
}
\label{fig:clump_fraction}
\end{figure*}

\subsection{Clump Frequency vs. Galaxy Properties} \label{sec:clumpy_demo}

We examine how the average number of clumps per galaxy ($\langle N_{\mathrm{clumps}} \rangle$) varies with redshift and structural parameters of the host galaxy. Figure~\ref{fig:clump_fraction} shows trends as a function of redshift, S\'ersic index, and Gini coefficient \citep{lotz_new_2004}. In addition to these morphological parameters, we also examine clump frequency as a function of specific star formation rate (sSFR) and stellar mass. We find that more massive galaxies tend to host more clumps, while sSFR shows only a weak correlation. To account for this, the figure also includes trends separately for low- and high-mass galaxies.

As Figure~\ref{fig:clump_fraction} shows, clump frequency increases toward $z \sim 2$, coinciding with the epoch of cosmic noon when galaxies are most actively forming stars and dynamically unstable. This trend is consistent with models in which high gas fractions, turbulence, and violent disk instabilities promote fragmentation into massive clumps \citep[e.g.,][]{dekel_cold_2009, ceverino_high-redshift_2010}. At higher redshifts ($z > 4$), clump counts decline, likely due to a combination of lower surface brightness in extended disks, increased incompleteness, and potentially smoother intrinsic morphologies. As recently shown by \citet{rinaldi_beyond_2025}, some high-redshift galaxies that resemble compact or low-surface-brightness systems may in fact be early analogs of cosmic-noon disks, with their outer structure falling below current detection limits. These redshift trends are broadly consistent across both low- and high-mass subsamples.

Host morphology, as traced by S\'ersic index $n$, shows a weaker correlation with clumpiness. Disk-like galaxies ($n \lesssim 2$) host more clumps on average than bulge-dominated systems ($n \gtrsim 4$), but the trend flattens at intermediate values. This suggests that single-component S\'ersic fits may not fully capture the structural diversity of clumpy systems, especially in irregular or interacting galaxies.

In contrast, the Gini coefficient ($G$), a non-parametric measure of flux concentration, correlates strongly with clump number. Galaxies with $G \gtrsim 0.7$ contain up to $\sim$4 to 7 clumps on average, while those with $G \lesssim 0.5$ host fewer than two. We focus on Gini because it is less sensitive to noise and segmentation uncertainties than other non-parametric indicators such as $M_{20}$ or asymmetry. Although high Gini values are often associated with central concentration, in our sample they more commonly reflect spatially distinct bright regions rather than smooth bulges. Visual inspection confirms that many high-Gini galaxies exhibit fragmented, asymmetric light distributions, not compact spheroids. Also, the observed trends with S\'ersic index and Gini coefficient are consistent across both stellar mass bins.

To separate redshift-driven effects from intrinsic structure, we analyze clump counts at fixed S\'ersic index and redshift (Appendix~\ref{sec:app_gini_test}). The positive trend with Gini persists in all bins, while the clump--redshift relation largely reflects the underlying evolution of Gini itself. This suggests that clumpiness is more directly tied to internal structural diversity than to redshift alone. Among commonly used structural metrics, the Gini coefficient emerges as the most reliable predictor of substructure richness in high-redshift galaxies. \added{In addition, to verify that this correlation is not driven by a mathematical artifact, we performed a mock test in Appendix~\ref{sec:app_gini_test} using synthetic galaxies composed of multiple identical Gaussian clumps at fixed total flux. The resulting Gini coefficients show no systematic dependence on clump number, confirming that the observed trend with Gini reflects intrinsic substructure contrast rather than pixel flux inequality.}

\subsection{Physical Properties of \added{Clumps}}
\label{sec:sed_properties}

\begin{figure*}[!ht]
\centering
\includegraphics[width=\linewidth]{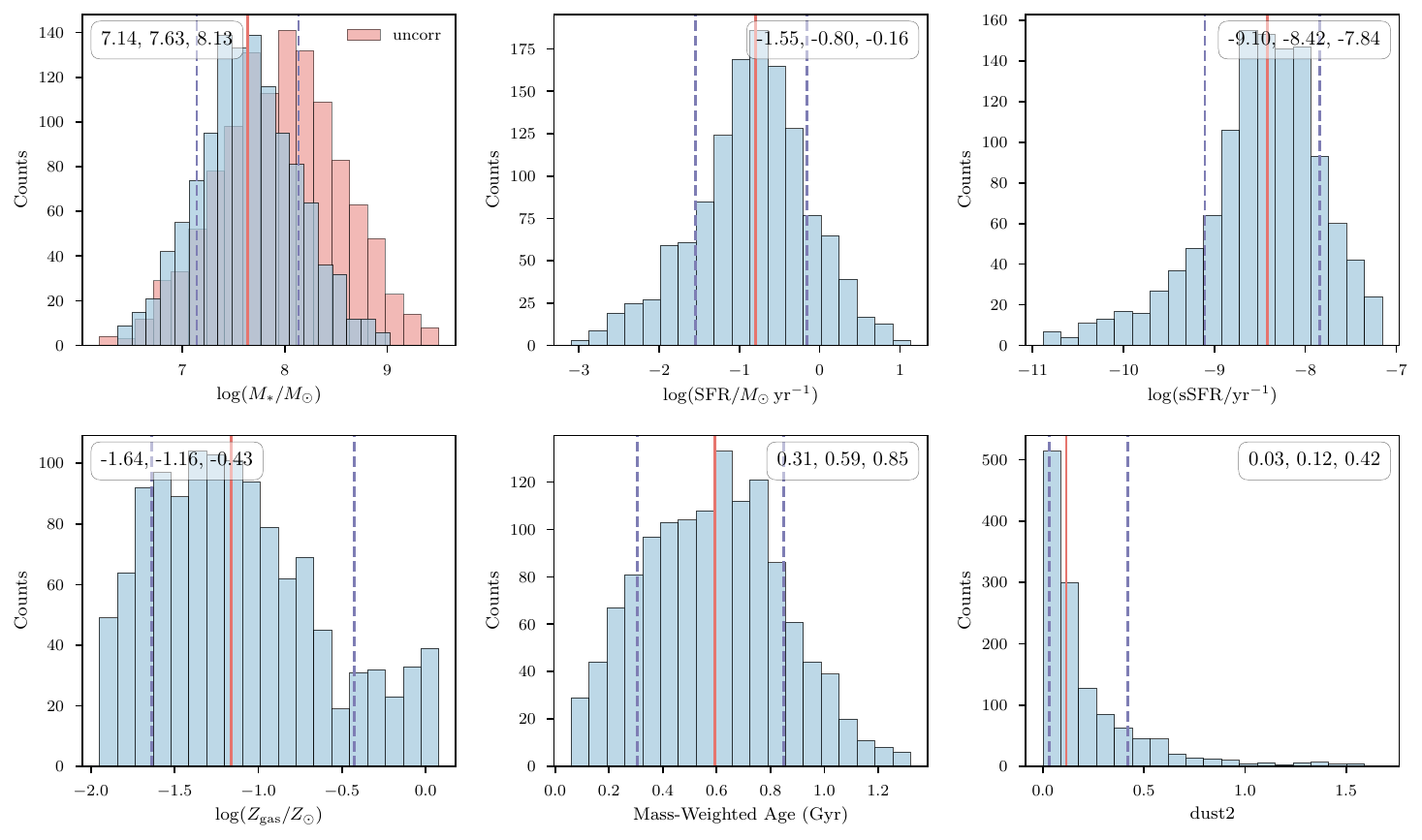}
\caption{
Distributions of physical properties derived from SED fitting of \added{clumps}.
Shown are (top row, left to right) stellar mass, star formation rate (SFR), and specific star formation rate (sSFR), and (bottom row) gas-phase metallicity, mass-weighted age, and dust attenuation parameter (\texttt{dust2} from \texttt{Prospector}, corresponding to the diffuse dust V-band optical depth). Vertical lines mark the 16th, 50th, and 84th percentiles of each distribution. In the stellar mass panel, the red histogram shows uncorrected values, while the blue histogram accounts for the fraction of flux in the residual image relative to the original image within the segmented clump regions. \added{Clumps} span a wide range of gas-phase metallicities and mass-weighted ages, indicating diverse stellar populations and evolutionary states. They are typically actively star forming, with a median $\log(\mathrm{sSFR}/\mathrm{yr}^{-1}) = -8.42$, and exhibit low to moderate dust attenuation. We caution that metallicity and age estimates are subject to significant uncertainties due to host light contamination and degeneracies in SED fitting, and should therefore be interpreted with care.
}
\label{fig:sed_properties}
\end{figure*}

We derive stellar population properties for \added{clumps} using multi-band aperture photometry and SED fitting. The size of each clump is measured in the NIRCam band that most closely corresponds to rest-frame 5000\,\AA\ (typically between F150W and F356W; see Section~\ref{sec:data}). For SED fitting, fluxes are extracted in all available NIRCam filters (F070W, F090W, F115W, F150W, F182M, F200W, F210M, F250M, F277W, F300M, F335M, F356W, F410M, and F444W), using images PSF-matched to the F444W resolution. The segmentation masks derived from residual detection are applied consistently across all bands to define the photometric apertures. These fluxes are fitted with the \texttt{Prospector} code \citep{johnson_stellar_2021}. While the photometry is not deblended from the host galaxy light, clumps in high-redshift disks typically exhibit similar stellar populations to their surroundings \citep{kalita_near-ir_2025}, so host contamination is not expected to strongly bias the inferred stellar mass and star formation rates. However, we caution against over-interpreting other derived parameters (such as age or gas-phase metallicity) since fully separating clump and host light is not feasible with the current data.

The SED fitting follows the \texttt{Prospector-$\alpha$} framework \citep{leja_how_2019}, adopting a Chabrier initial mass function (IMF; \citealt{chabrier_galactic_2003}) and a non-parametric star formation history with 7 age bins and a continuity prior that allows smooth variations in star formation over time. The models include both stellar and nebular emission and fit for stellar mass, age, gas-phase metallicity, dust attenuation, and star formation rate. To accelerate the fitting process, we employ the artificial neural network (ANN) emulator \texttt{Parrot} \citep{mathews_as_2023}, which reproduces full \texttt{Prospector} results with typical deviations of $\lesssim$0.15\,dex in key physical parameters while reducing computation time by more than an order of magnitude (also see e.g., \citealp{alberts_smiles_2024,zhu_higher_2025}).

Figure~\ref{fig:sed_properties} summarizes the distributions of stellar mass, gas-phase metallicity, mass-weighted stellar age, star formation rate (averaged over the past 30 Myr), and dust attenuation (\texttt{dust2}). Clump masses span $\log(M_\star/M_\odot) \sim 7$-8, comparable to dwarf galaxies or massive star-forming regions in turbulent disks \citep[e.g.,][]{zanella_contribution_2019}. Flux-fraction-corrected estimates (blue histogram) are systematically lower than uncorrected values (red), consistent with modest host contamination. Gas-phase metallicity values peak near $\log(Z/Z_\odot) \sim -1.1$ but extend up to solar, within the adopted flat prior of $-2 < \log(Z/Z_\odot) < 0.5$, indicating that clumps are typically embedded in enriched environments but may also include younger, lower-metallicity regions. Mass-weighted ages have a broad distribution, with a median of $\sim$0.5 Gyr and most clumps younger than 1 Gyr. However, these values may reflect the cumulative star formation history rather than recent bursts. Smoothly rising SFHs or weak age constraints from photometry can yield intermediate ages even for clumps dominated by young stars. Moreover, without a clean decomposition of clump light from the underlying host galaxy, the derived ages may be biased by older stellar populations in the surrounding disk. We therefore caution against interpreting mass-weighted age as direct evidence of recent formation.

Clump SFRs span a broad range with a median of $\sim 0.2\,M_\odot\,\mathrm{yr}^{-1}$, consistent with active star formation. Specific star formation rates range from $\log({\rm sSFR}/{\rm yr}^{-1}) = -9.10$ to $-7.84$, with a median of $-8.42$, further supporting the view that these substructures are actively forming stars. We caution that both SFR and sSFR values integrate the emission from the full segmented region and may include modest host-galaxy light contamination \citep[see also][]{kalita_near-ir_2025}. Dust attenuation is generally low to modest, with \texttt{dust2} peaking around 0.1 (corresponding to $A_V \approx 0.11$\,mag) and few clumps exceeding 0.5 ($A_V \approx 0.54$\,mag), in line with expectations for compact star-forming regions in low-mass, gas-rich systems where feedback may efficiently clear surrounding material \citep[e.g.,][]{newman_shocked_2012}. Together, these results suggest that the \added{clumps} in JADES GOODS--South galaxies are actively star-forming, low-mass, and typically have modest dust attenuation, broadly consistent with in-situ formation in unstable, high-redshift disks.

\subsection{Size--Mass Relation and Redshift Trends}
\label{sec:size_mass}

\begin{figure}[!ht]
\centering
\includegraphics[width=\columnwidth]{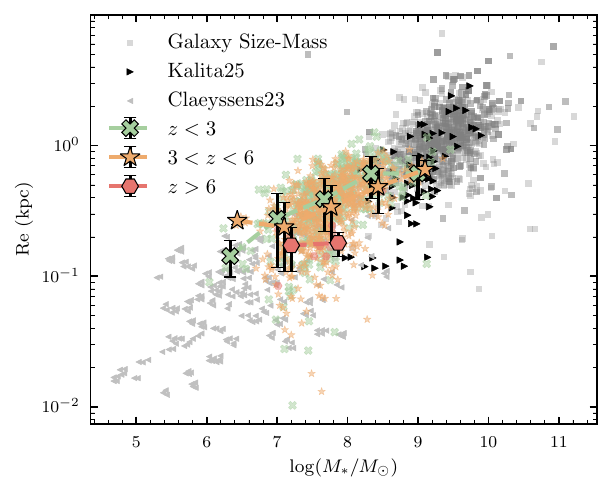}
\caption{
Size--mass relation of \added{clumps} in \textsc{JADES} GOODS--South, shown in three redshift \added{ranges}: $z<3$ (green), $3<z<6$ (orange), and $z>6$ (red).
Semi-transparent points show individual clumps; large symbols mark the mean $R_{\rm e}$ in each stellar mass bin, with error bars indicating the 16th--84th percentile range.
For comparison, gray squares show the host-galaxy size--mass relation measured from the same sample, gray triangles represent lensed clumps from \citet{claeyssens_star_2023} at $z \sim 1$--5, and black triangles denote NIRCam-resolved clumps from \citet{kalita_clumps_2025} at $1 < z < 2$.
The host size--mass relation measured here is broadly consistent with other recent JADES results \citep[e.g.,][]{danhaive_beyond_2025} and measurements from the literature \citep[e.g.,][]{allen_galaxy_2025, miller_jwst_2025, yang_cosmos-web_2025}.
Overall, the JADES clumps occupy an intermediate regime between compact lensed star-forming knots and their host galaxies, connecting small-scale substructures to the global size--mass relation across cosmic time.
}
\label{fig:size_mass}
\end{figure}

\begin{figure*}[!ht]
\centering
\includegraphics[width=0.32\linewidth]{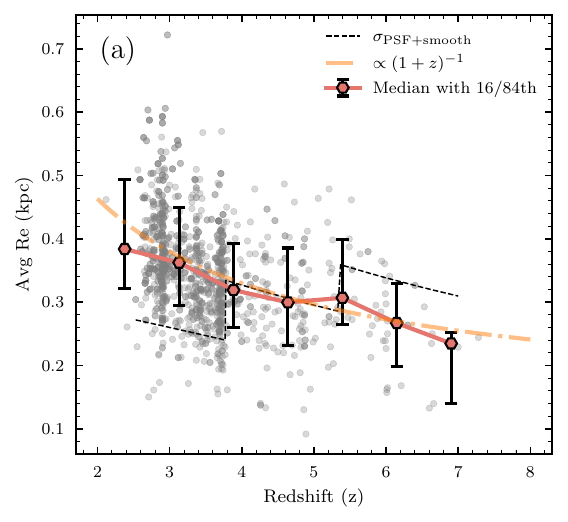}
\includegraphics[width=0.32\linewidth]{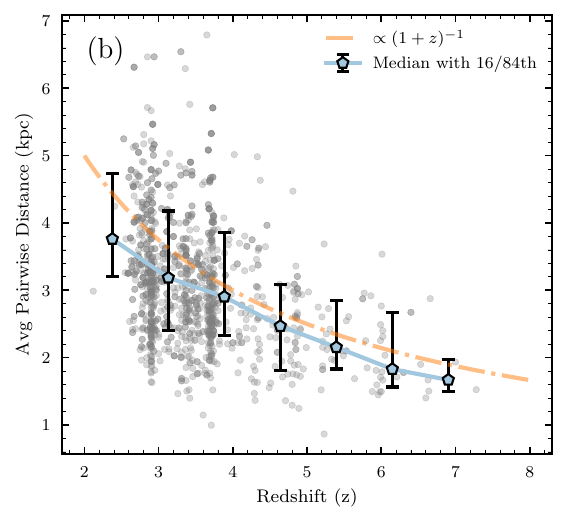}
\includegraphics[width=0.32\linewidth]{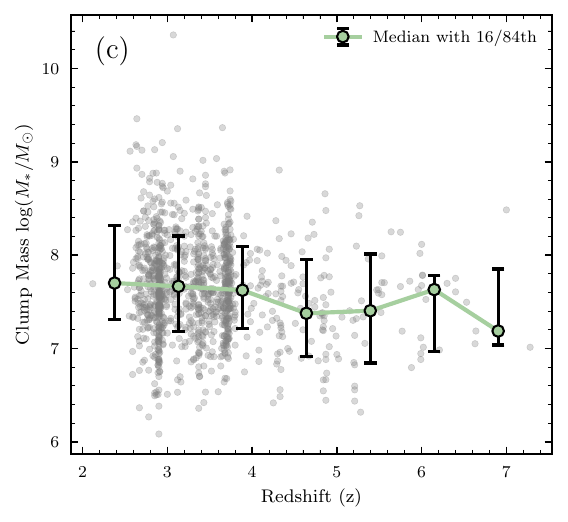}
\caption{
Redshift evolution of structural properties of \added{clumps} in \textsc{JADES} GOODS--South galaxies. 
\textbf{(a)} Average clump effective radius ($R_{\rm e}$) after subtracting the blending scale in quadrature, defined as $\sigma_{\rm PSF+smooth} \approx \sqrt{\sigma_{\rm PSF}^2 + (1\,\mathrm{pix})^2}$. 
The dashed line marks this approximate resolution limit, dominated by the PSF. 
\textbf{(b)} Average pairwise physical distance between clumps within individual galaxies. In both panels (a) and (b), the dotted-dashed orange line shows the rescaled galaxy size evolution ($\propto (1+z)^{-1}$; see \citealp{danhaive_beyond_2025} and references therein). 
\textbf{(c)} Stellar mass of individual clumps. 
Gray points show individual measurements, and colored lines indicate the median trend, with error bars representing the 16th to 84th percentile range in each redshift bin. 
Both the typical clump size and spacing decrease toward higher redshift, consistent with expectations from inside-out galaxy growth and increasing gas surface density at early times. 
In contrast, the median clump stellar mass shows no statistically significant evolution with redshift (Spearman $p > 0.05$), although the scatter increases toward cosmic noon.
}
\label{fig:z_evolution}
\end{figure*}

We examine how the effective radius ($R_{\rm e}$) of \added{clumps} scales with stellar mass and redshift. Figures~\ref{fig:size_mass} and~\ref{fig:z_evolution} present complementary views of these trends. Clump sizes are measured from the residual-segmentation masks (Section~\ref{sec:method}) as the \emph{circular} half-light radius: for each clump, we rank pixels by their Euclidean distance from the clump centroid, accumulate flux, and define $R_{\rm e}$ at 50\% of the enclosed light. These radii are measured from PSF-blurred segmentation maps and converted to physical units using the angular-diameter distance at the host redshift. To account for PSF broadening and minimal pixel-level blending, we apply an approximate deblending correction by subtracting the effective smoothing scale, $\sqrt{\sigma_{\rm PSF}^2 + (1\,\mathrm{pix})^2}$, in quadrature. The resulting $R_{\rm e}$ values are thus corrected for the dominant beam smearing. However, residual biases may persist (particularly for compact or irregular clumps) since the PSF also influences the segmentation process and internal flux gradients. These effects tend to overestimate intrinsic sizes. Still, within narrow redshift ranges where PSF and depth are similar, the relative size distribution remains robust.

Figure~\ref{fig:size_mass} shows the size--mass relation of \added{clumps} in three redshift bins, along with the host galaxy size--mass relation measured from the same JADES sample (gray squares) for comparison. Both clump and galaxy sizes are measured in the NIRCam wide-band filter corresponding most closely to rest-frame $0.5\,\mu{\rm m}$ at each galaxy's redshift, ensuring consistent wavelength sampling across the full redshift range. Each clump is shown individually (semi-transparent points), with binned averages indicated by large symbols and vertical error bars marking the 16th-84th percentile range. At all redshifts, clump size increases with stellar mass, consistent with expectations for self-gravitating, virialized structures. The host galaxy size--mass relation measured here is broadly consistent with recent JADES measurements \citep[e.g.,][]{danhaive_beyond_2025} and measurements from literature \citep[e.g.,][]{allen_galaxy_2025,miller_jwst_2025,yang_cosmos-web_2025}.

The normalization of this relation evolves with redshift: at fixed stellar mass, clumps are larger at later times. This trend likely reflects a combination of inside-out disk growth and declining ISM turbulence, which shifts the characteristic fragmentation scale to larger physical sizes over time \citep[e.g.,][]{ceverino_rotational_2012, tamburello_clumpy_2017}. The systematic increase in spatial extent toward lower redshift mirrors the evolution of the global galaxy size--mass relation.

For context, gray squares denote the \textit{host galaxy} size--mass relation measured from the same JADES sample, while gray triangles show lensed clumps from \citet{claeyssens_star_2023} at $z\sim1$-5 and black triangles represent NIRCam-resolved clumps from \citet{kalita_clumps_2025} at $z=1.43$-1.74. The JADES clumps populate the intermediate regime between compact, lensed star-forming knots and their extended host galaxies, linking small-scale fragmentation within disks to the global buildup of galaxy structure across cosmic time.

To isolate redshift evolution independently of mass, we compute the median clump size in narrow redshift bins (Figure~\ref{fig:z_evolution}a). The typical $R_{\rm e}$ decreases from $\sim$0.4\,kpc at $z\sim2.5$ to $\sim$0.25\,kpc at $z\sim7$, with moderate scatter. The redshift evolution seen in clump sizes mirrors the well-known growth of galaxy sizes over cosmic time \citep[e.g.,][and references therein]{danhaive_beyond_2025}. This smooth evolution supports a scenario in which fragmentation occurs on progressively smaller spatial scales in the early Universe and shifts outward as galaxies grow and stabilize. 

We also measure the average pairwise physical distance between clumps within each galaxy (Figure~\ref{fig:z_evolution}b). The mean spacing declines from $\sim$4\,kpc at $z\sim2.5$ to $\sim$2\,kpc at $z\sim7$, mirroring the size evolution and indicating more compact clump configurations at early times. This trend parallels the redshift evolution of global galaxy sizes and is consistent with inside-out disk growth and hierarchical buildup \citep[e.g.,][]{baker_core_2025}. In contrast, the median clump stellar mass (Figure~\ref{fig:z_evolution}c) shows no statistically significant evolution with redshift (Spearman $p>0.05$), although a larger scatter is present, suggesting diverse formation conditions among individual clumps.

\subsection{Clump Mass Function}
\label{sec:mass_function}

\begin{figure}[!ht]
\centering
\includegraphics[width=\columnwidth]{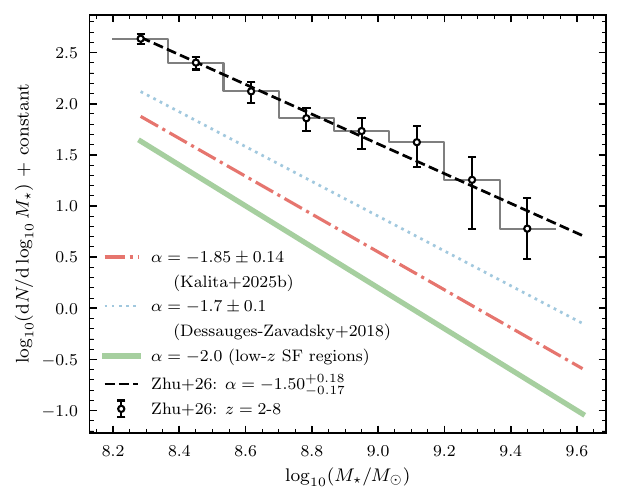}
\caption{
Stellar mass function of \added{clumps} identified in this work (\added{$2<z<8$}). 
Black circles show our mass function with bootstrap $1\sigma$ uncertainties. 
The dashed black line shows the best-fit power-law relation, $\mathrm{d}N/\mathrm{d}M_\star \propto M_\star^{\alpha}$, with a slope of $\alpha = -1.50^{+0.18}_{-0.17}$. 
For comparison, colored lines show reference slopes from previous studies: 
$\alpha = -1.85 \pm 0.14$ from the model-based clump analysis of \citet{kalita_clumps_2025}, 
$\alpha = -1.7 \pm 0.1$ from lensed high-redshift galaxies \citep{dessauges-zavadsky_first_2018}, 
and $\alpha = -2.0$ for nearby star-forming regions \citep[e.g.,][]{ elmegreen_hierarchical_2006}. 
The JADES clump mass function is shallower than the self-similar prediction and lies between the hierarchical and top-heavy regimes, indicating that high-redshift clumps form in a moderately top-heavy, merger-influenced environment, consistent with the measurements at $1<z<2$ in \citet{kalita_near-ir_2025}.
}
\label{fig:mass_function}
\end{figure}

We measure the stellar mass function of \added{clumps} over the redshift range $z=2$--\added{8}. Figure~\ref{fig:mass_function} shows the \emph{differential} mass function computed per dex, $\phi \equiv \mathrm{d}N/\mathrm{d}\log_{10} M_\star$, for clumps with $\log(M_\star/M_\odot) > 8.2$. We adopt equal-width bins in $\log_{10} M_\star$ between 8.2 and 9.7 (bin width $\simeq0.15$\,dex, similar to \citealt{kalita_near-ir_2025}). In each bin, we compute $\phi_i = N_i / \Delta\log_{10} M_\star$ and plot $\log_{10}\phi$ versus $\log_{10} M_\star$; bins with zero counts are excluded from the fit. The dashed line indicates a linear fit over the completeness-limited range $\log(M_\star/M_\odot)=8.2$-9.6. 

The resulting clump mass function is well described by a single power law with $\alpha = -1.50^{+0.18}_{-0.17}$. This slope is fully consistent with the canonical hierarchical prediction of $\alpha \simeq -1.5$ \citep[e.g.,][]{hopkins_general_2013}, although is shallower than the self-similar expectation ($\alpha=-2$), and is steeper than a top-heavy case ($\alpha=-1$). For comparison, Figure~\ref{fig:mass_function} also shows reference slopes from previous studies: $\alpha = -1.85 \pm 0.14$ from the model-based clump analysis of \citet{kalita_clumps_2025}, and $\alpha = -1.7 \pm 0.1$ from lensed high-redshift galaxies \citep{dessauges-zavadsky_first_2018}. Overall, the JADES clump mass function aligns with hierarchical fragmentation and is broadly consistent with measurements of $\alpha=-1.50\pm 0.14$ at $1<z<2$ in \citet{kalita_near-ir_2025}. \added{We have verified that this slope remains stable against reasonable variations in mass threshold and redshift range, and that residual incompleteness primarily affects the normalization rather than the differential slope (Appendix~\ref{sec:app_completeness}).}

Our result favors a scenario in which clumps form through a combination of stochastic fragmentation and regulated collapse, shaped by turbulence, disk instability, and local gas conditions, rather than a purely merger-driven or self-similar process. This interpretation aligns with high-resolution simulations of unstable disks at high redshift, including those from the VDI and FIRE frameworks, which typically predict clump mass-function slopes in the range $\alpha \sim -1.3$ to $-1.5$ \citep[e.g.,][]{ceverino_rotational_2012, tamburello_clumpy_2017, mandelker_giant_2017}.

\subsection{Radial Distribution and Structural Variation}
\label{sec:radial}

\begin{figure*}[!ht]
\centering
\includegraphics[width=0.6\linewidth]{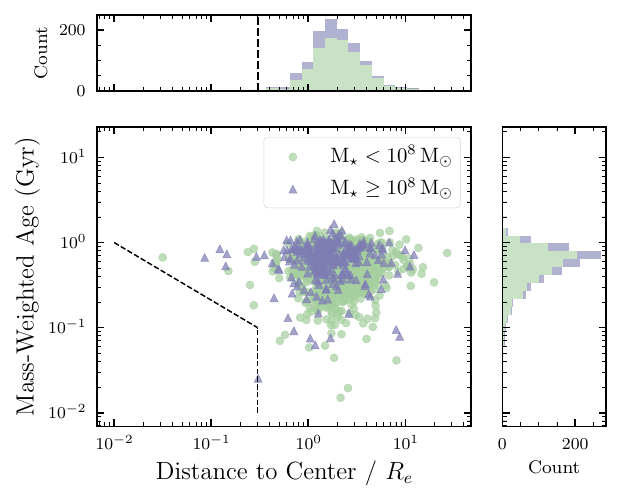}
\caption{
Radial distribution of \added{clumps} in JADES GOODS--South galaxies. 
The horizontal axis shows galactocentric distance normalized by the host galaxy's effective radius ($R/R_{\rm e}$), and the vertical axis shows clump age.
The dashed region highlights a central ``deadzone'', a relative deficit of young clumps near galaxy centers ($R/R_{\rm e} \lesssim 0.3$), which may reflect inward migration of older clumps, disruption by feedback, or detection incompleteness in crowded central regions.
The observed trend is consistent with an \textit{outside-in} clump formation scenario, where clumps form in outer disks and migrate inward over time.
}
\label{fig:radial_profiles}
\end{figure*}

We examine how clump properties vary with position inside galaxies, focusing on stellar age and structural scaling as a function of galactocentric distance normalized by the host effective radius ($R/R_{\rm e}$). This analysis tests for signatures of clump migration, inside-out growth, or environment-driven evolution. Here, $R$ denotes the projected circular distance between the clump and the galaxy center in the image plane. We do not apply inclination or ellipticity corrections, since reliable axis-ratio measurements are unavailable for many high-redshift systems and the projection effects are expected to average out statistically across the large sample.

Figure~\ref{fig:radial_profiles} shows the distribution of clump mass-weighted stellar age versus $R/R_{\rm e}$, with clumps split into low-mass ($M_\star < 10^8\,M_\odot$; green circles) and high-mass ($M_\star \geq 10^8\,M_\odot$; purple triangles) subsets. A central deficit of young clumps is evident within $R/R_{\rm e} \lesssim 0.3$, marked by the dashed region. This \added{apparent} ``deadzone'' may reflect (1) inward migration of older clumps from the disk, (2) disruption of young clumps by strong central feedback, or (3) reduced completeness in crowded central regions. 
\added{To quantify the potential impact of incompleteness, we estimate an upper limit on the intrinsic fraction of young clumps in the central region ($R/R_{\rm e} < 0.3$), defining ``young'' as clumps with mass-weighted age below the sample median (MWA $<0.59$\,Gyr). We observe $N_{\rm tot,obs}=12$ clumps in this region, of which $N_{\rm young,obs}=4$, corresponding to $f_{\rm young,center}=0.33$. Using the injection-recovery completeness at small radii (Appendix~\ref{sec:app_completeness}), the recovery fraction is $C\simeq0.51$ (weighted average) or $C\simeq0.41$ (conservative value). Correcting for incompleteness implies an intrinsic central clump population of $N_{\rm tot,corr}\sim24$--29. In the extreme scenario that all missed clumps were young, the upper limit on the intrinsic young fraction would be $f_{\rm young,center}<0.66$ (weighted) or $<0.73$ (conservative). For comparison, in the outer region ($0.3\le R/R_{\rm e}<2.5$) the observed young fraction is $f_{\rm young,outer}=0.45$ (352/790). These estimates show that incompleteness alone cannot be ruled out as a contributor to the central trend, but the qualitative radial pattern remains consistent with structural evolution.}

The observed age-radius trend is consistent with an \textit{outside-in} clump formation scenario, where clumps form via gravitational instability at larger radii and migrate inward over time \citep[e.g.,][]{dekel_clump_2022, ceverino_rotational_2012}. This naturally explains the buildup of older clumps near the center and younger ones in the outskirts, as also seen in simulations and recent JWST studies \citep{zanella_contribution_2019, kalita_clumps_2025}.

\begin{figure*}[!ht]
\centering
\includegraphics[width=0.85\linewidth]{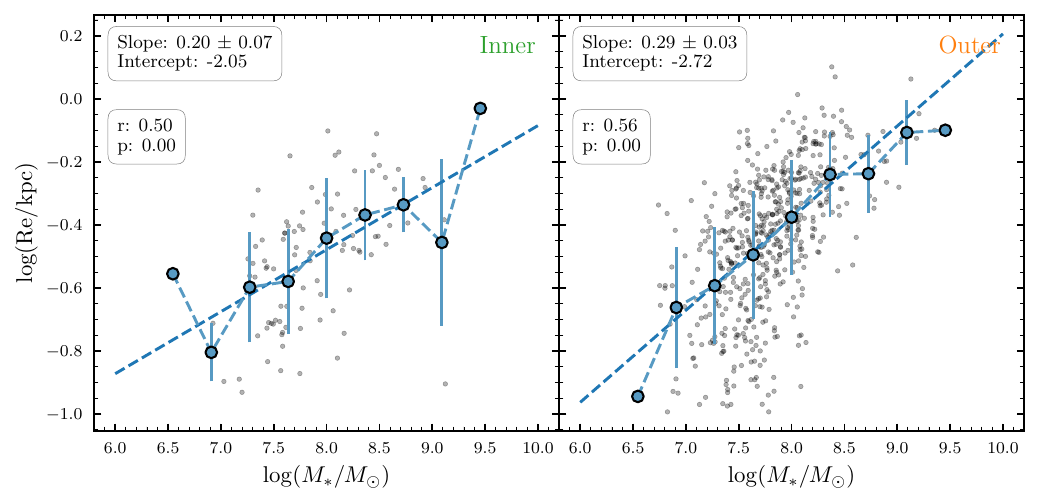}
\caption{
size--mass relation of \added{clumps} in \textsc{JADES} GOODS--South, separated by normalized galactocentric distance.
The left and right panels show clumps in the inner ($R/R_{\rm e}<1$) and outer ($1<R/R_{\rm e}<2.5$) regions, respectively, combining all redshifts.
Gray points represent individual clumps, blue circles and error bars indicate binned mean with 16th-84th percentile scatter, and dashed blue lines show the best-fit linear relations.
Outer clumps follow a steeper slope ($0.29\pm0.03$), approaching the virial expectation of $R_{\rm e}\propto M_*^{1/3}$, while inner clumps show a shallower scaling ($0.20\pm0.07$), implying higher surface densities or stronger dynamical disturbance.
}
\label{fig:radial_smr}
\end{figure*}

However, as mentioned earlier, the age measurements for these clumps are less reliable \added{and potentially vulnerable to completeness effects}; structural diagnostics provide a more empirical and model-independent view of clump evolution. To test for structural evolution, we measure the clump size--mass relation separately in the inner ($R/R_{\rm e}<1$) and outer ($R/R_{\rm e}>1$) regions of galaxies, as shown in Figure~\ref{fig:radial_smr}. When combining all redshifts, outer clumps exhibit a steeper size--mass slope ($R_{\rm e} \propto M^{0.29\pm0.03}$), approaching the virial relation expected for self-gravitating systems ($R \propto M^{1/3}$). In contrast, inner clumps follow a shallower slope ($R_{\rm e} \propto M^{0.20\pm0.07}$), consistent with tidally compressed or dynamically disturbed structures, or with regions affected by strong feedback and shear \citep[e.g.,][]{liu_wisdom_2022}. Such a slope also agrees with the size--mass relation for galaxies across cosmic time \citep[e.g.,][]{danhaive_beyond_2025}. A bootstrap analysis confirms that the slope difference between inner and outer clumps ($\Delta=0.09\pm0.07$, $p\simeq0.03$) is statistically significant, suggesting that outer clumps are more virialized and self-regulated, whereas inner clumps are more compact and may be in the process of migration or coalescence toward the galactic center.

This radial dependence supports a migration-driven evolution of clump properties. Clumps formed in outer disks may retain their initial virialized structure, while those that migrate inward experience compression or disruption, flattening their size--mass scaling. The consistency between these structural trends and the age-radius gradient strengthens the case for clump migration as a driver of bulge growth and morphological transformation.

Although detection biases near galaxy centers may impact these measurements, our injection-recovery tests (Figure~\ref{fig:completeness}) indicate that completeness
\added{declines toward the very center but remains moderate above our adopted thresholds.}
\added{In addition, restricting the analysis to clumps with high residual-to-total flux ratios yields consistent radial trends (Appendix~\ref{sec:app_host_contam}), suggesting that host-light contamination does not drive the observed gradients.}
These results suggest that clumps evolve structurally as they move through the galaxy, contributing to central mass buildup over cosmic time.

\section{Discussion} \label{sec:discussion}

The \added{clumps} identified in this study provide new insights into the formation and internal evolution of high-redshift galaxies. Their physical properties are broadly consistent with in-situ formation in gas-rich, turbulent disks. Such environments, common at $z \sim 2$-4, are prone to gravitational fragmentation under the Toomre instability criterion, as seen in simulations of VDI \citep[e.g.,][]{dekel_clump_2022, ceverino_high-redshift_2010, inoue_non-linear_2016}.

The strong redshift evolution in clump incidence, rising toward $z \sim 2$, along with the tight correlation between clump number and Gini coefficient and the weak dependence on S\'ersic index, supports a scenario in which internal structural diversity, rather than overall galaxy morphology, governs substructure formation. The clump mass function follows a power-law slope of $\alpha = -1.50_{-0.17}^{+0.19}$, consistent with predictions from high-resolution simulations of unstable disks \citep[e.g.,][]{ceverino_rotational_2012, mandelker_giant_2017}. This intermediate slope suggests that clump formation is not purely hierarchical or stochastic, but instead reflects a regulated balance between turbulence, self-gravity, and feedback.

A particularly striking result is the deficit of young clumps at $R/R_{\rm e} \lesssim 0.3$, accompanied by older ages and structural compression in central regions. This “deadzone” is qualitatively consistent with inward clump migration, where massive clumps lose angular momentum and sink toward the nucleus via dynamical friction or torques \citep{bournaud_rapid_2007, inoue_natures_2012}. The radial gradient in clump age and size--mass scaling provides complementary evidence for this process. While detection biases near galaxy centers may contribute, injection-recovery tests suggest that incompleteness alone cannot account for the observed trends (Appendix~\ref{sec:app_completeness}).

We also observe coherent structural trends: clump size increases with stellar mass, and both sizes and pairwise clump separations grow toward lower redshifts. Outer clumps exhibit steeper size--mass slopes approaching virial expectations ($\sim 1/3$), while inner clumps show flatter scaling. These patterns point to differing evolutionary pathways depending on clump environment and galactocentric distance.

Not all clumps survive to migrate inward. Low-mass or compact clumps may be rapidly disrupted by feedback, supernova-driven winds, or shear from differential rotation \citep{genel_short-lived_2012}. The shallow slope of the clump mass function implies that a few massive clumps dominate the total substructure mass budget, while many lower-mass clumps are likely short-lived. Future observations, including resolved gas kinematics and emission-line diagnostics from NIRSpec, will help distinguish between bound and unbound clumps, and clarify the impact of feedback on clump survival.

Our analysis is subject to several limitations. The residual-based detection method depends on single-component S\'ersic subtraction and may misidentify asymmetric light in irregular systems as clumps. Photometry is measured from residual-defined apertures that include some host light, and we do not model clump morphologies directly. While tests from \citet{kalita_clumps_2025} and our own injection simulations suggest these effects are minor, a more sophisticated approach involving multi-component fitting and forward modeling would reduce systematic uncertainties. In addition, residual-based methods are intrinsically more sensitive to asymmetric or off-center features, potentially favoring the detection of peripheral substructures while underestimating compact or centrally embedded clumps. This bias could modestly enhance the observed radial gradients in clump incidence and structure, although completeness tests (Appendix~\ref{sec:app_completeness}) indicate that such effects are secondary to genuine physical trends.

Despite these caveats, our results demonstrate the power of residual-based analysis for large, unlensed galaxy samples. The methods presented here enable consistent clump identification across $\sim$3600 galaxies over $z = 2$--8, revealing statistically robust trends in clump incidence, structure, and evolution. Incorporating future spectroscopic and kinematic datasets will further constrain clump lifetimes, formation channels, and their role in building galaxy structure over cosmic time.

\section{Summary} \label{sec:summary}

We present a statistical study of \added{clumps} in $\sim$3600 galaxies from JWST/NIRCam imaging in the JADES GOODS--South field, spanning $z = 2$--8. Our main findings are:

\begin{enumerate}
    \item Clumps are identified as residual features after S\'ersic profile subtraction. They are typically low-mass ($\log M_\star/M_\odot \sim 7$--8), actively star-forming (sSFR $\sim 10^{-9}$--$10^{-7}$ yr$^{-1}$), with modest dust attenuation.

    \item Clump frequency increases toward $z \sim 2$ and correlates strongly with Gini coefficient, but shows weaker dependence on S\'ersic index. This suggests that internal structural complexity, rather than overall galaxy morphology, is the primary driver of clump formation.

    \item Clumps follow a positive size--mass relation, and both their typical size and pairwise separation increase toward lower redshifts, consistent with inside-out disk growth.

    \item The clump stellar mass function follows a power law with slope $\alpha = -1.50_{-0.17}^{+0.19}$, in line with predictions from simulations of turbulent disk fragmentation.

    \item An \added{apparent} deficit of young clumps at $R/R_{\rm e} \lesssim 0.3$ suggests a central ``deadzone,'' possibly shaped by clump migration, disruption, or detection incompleteness.

    \item Clumps in outer disks exhibit steeper size--mass slopes and are generally younger, while inner clumps are smaller, older, and structurally flatter. This is broadly consistent with inward migration.
\end{enumerate}

These results support a scenario in which clumps originate from gravitational instabilities within gas-rich, turbulent disks and subsequently evolve as they migrate inward, contributing to bulge growth and the morphological transformation of their hosts. Looking ahead, combining JWST imaging with spatially resolved spectroscopy and ALMA observations will directly probe the gas, dust, and ionized components of individual clumps, enabling constraints on their lifetimes, feedback energetics, and molecular gas reservoirs. Together, these multiwavelength datasets will establish a comprehensive, empirical framework for how small-scale star-forming units assemble and shape galaxies across cosmic time.

\begin{acknowledgments}
YZ, MJR, ZJ, CC, DJE, BDJ, GHR, BR, and CNAW acknowledge support from the NIRCam Science Team contract to the University of Arizona, NAS5-02105. YZ is also supported by JWST Program \#6434. CC, DJE, and BR are also supported by JWST Program \#3215.
Support for program \#6434 and \#3215 was provided by NASA through a grant from the Space Telescope Science Institute, which is operated by the Association of Universities for Research in Astronomy, Inc., under NASA contract NAS 5-03127.
AJB acknowledges funding from the ``FirstGalaxies'' Advanced Grant from the European Research Council (ERC) under the European Union's Horizon 2020 research and innovation program (Grant agreement No. 789056).
A.L.D. thanks the University of Cambridge Harding Distinguished Postgraduate Scholars Programme and Technology Facilities Council (STFC) Center for Doctoral Training (CDT) in Data intensive science at the University of Cambridge (STFC grant number 2742605) for a PhD studentship.
ST acknowledges support by the Royal Society Research Grant G125142.
H\"U acknowledges funding by the European Union (ERC APEX, 101164796). Views and opinions expressed are however those of the authors only and do not necessarily reflect those of the European Union or the European Research Council Executive Agency. Neither the European Union nor the granting authority can be held responsible for them.

This work is based on observations made with the NASA/ESA/CSA James Webb Space Telescope. The data were obtained from the Mikulski Archive for Space Telescopes at the Space Telescope Science Institute, which is operated by the Association of Universities for Research in Astronomy, Inc., under NASA contract NAS 5-03127 for JWST. All the JWST data used in this paper can be found in MAST: \dataset[https://doi.org/10.17909/8tdj-8n28]{https://doi.org/10.17909/8tdj-8n28} \citep{JADES-data}. The authors acknowledge the FRESCO (PI: P.~Oesch) team for developing their observing program with a zero-exclusive-access period.

This material is based upon High Performance Computing (HPC) resources supported by the University of Arizona TRIF, UITS, and Research, Innovation, and Impact (RII) and maintained by the UArizona Research Technologies department. This project made use of lux supercomputer at UC Santa Cruz, funded by NSF MRI grant AST 1828315. 

We respectfully acknowledge the University of Arizona is on the land and territories of Indigenous peoples. Today, Arizona is home to 22 federally recognized tribes, with Tucson being home to the O'odham and the Yaqui. The university strives to build sustainable relationships with sovereign Native Nations and Indigenous communities through education offerings, partnerships, and community service.

This manuscript benefited from grammar checking and proofreading using ChatGPT \citep{openai_chatgpt_2024}.

\end{acknowledgments}


\vspace{5mm}
\facilities{JWST, MAST}

\software{
    {\tt Astropy} \citep{astropy_collaboration_astropy_2022},
    {\tt Matplotlib} \citep{hunter_matplotlib_2007},
    {\tt NumPy} \citep{van_der_walt_numpy_2011},
    {\tt scikit-image} \citep{van_der_walt_scikit-image_2014}
}

\appendix

\renewcommand{\thefigure}{A\arabic{figure}}
\setcounter{figure}{0}

\begin{figure*}[!ht]
\centering
\includegraphics[width=0.8\linewidth]{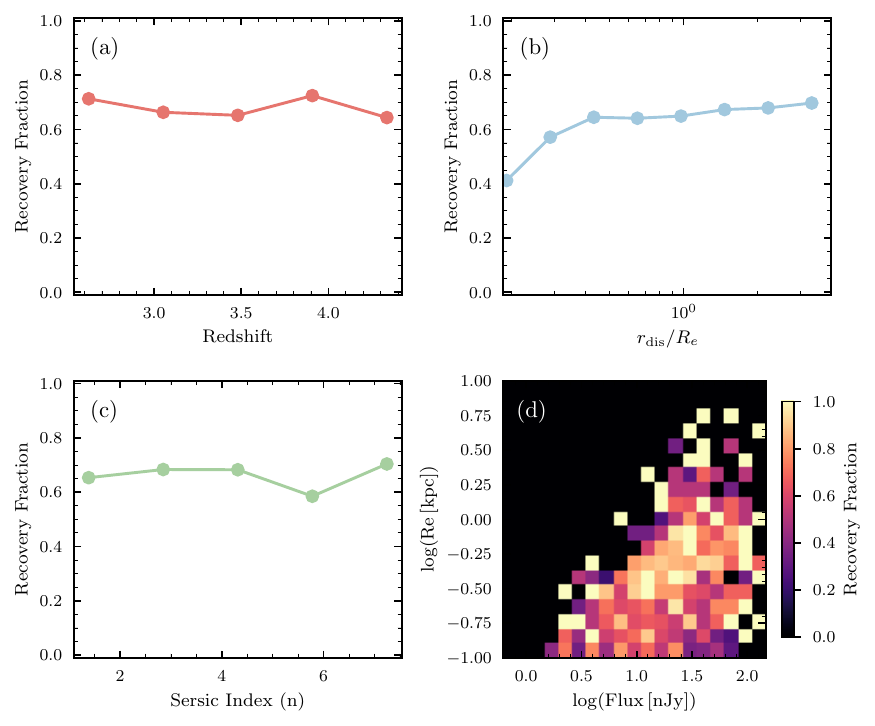}
\caption{
Detection completeness of \added{clumps} derived from injection-recovery simulations in JADES GOODS--South. 
\textbf{(a)} Recovery fraction as a function of redshift, showing mild variation over $z \sim 2.5$-4.2. 
\textbf{(b)} Recovery fraction as a function of galactocentric distance normalized by host galaxy effective radius ($r_{\rm dis}/R_{\rm e}$), showing a clear decline toward the inner regions of galaxies.
\textbf{(c)} Recovery fraction as a function of host galaxy S\'ersic index, indicating weak dependence on overall morphology. 
\textbf{(d)} 2D completeness map as a function of clump flux and effective radius, revealing the highest recovery rates for bright and compact clumps. 
These trends indicate that incompleteness-particularly in the central regions and for faint, extended clumps-may contribute to the apparent deficit of young clumps near galaxy centers.
}
\label{fig:completeness}
\end{figure*}

\section{Clump Detection Completeness}
\label{sec:app_completeness}

To quantify detection biases, we perform injection-recovery simulations to assess the completeness of clump identification as a function of physical and morphological parameters. Artificial clumps modeled as Gaussian light profiles with known fluxes and effective radii were inserted into real residual images of representative JADES GOODS--South galaxies, spanning a range of redshifts, S\'ersic indices, and sizes. We then applied the standard detection pipeline (Section~\ref{sec:method}) to determine recovery rates.
As noted by e.g., \citet{rinaldi_beyond_2025}, high-redshift galaxies tend to be intrinsically smaller, which could therefore imply proportionally smaller clumps that are harder to detect at fixed surface-brightness limits. This possible coupling between galaxy size and clump detectability is not modeled in our simulations and remains a caveat for interpreting completeness at the highest redshifts.

Figure~\ref{fig:completeness} summarizes the results. In panel (a), completeness is shown as a function of redshift. Recovery is relatively stable across $z \sim 2.5$--4.2, with a mild decline at higher redshifts, likely reflecting reduced surface brightness and increased noise.

Figure~\ref{fig:completeness} (b) shows completeness versus normalized galactocentric distance ($r_{\rm dis}/R_{\rm e}$), revealing a clear drop in recovery toward galaxy centers. This is expected due to higher background levels, residual subtraction artifacts, and segmentation confusion in crowded regions. These effects are important when interpreting the deficit of young clumps at small radii (Figure~\ref{fig:radial_profiles}).

In Figure~\ref{fig:completeness} (c), we examine completeness as a function of host galaxy S\'ersic index. The results show little dependence on morphology, suggesting that our method is not strongly biased toward disk- or bulge-dominated hosts.

Finally, Figure~\ref{fig:completeness} (d) presents a 2D completeness map as a function of clump flux and size. As expected, bright and compact clumps are recovered most efficiently, while faint or extended clumps are less likely to be detected. The majority of clumps in our sample fall in the high-completeness regime.

Overall, these tests confirm that our detection method is robust across a wide range of clump and host properties. However, completeness declines for small, faint clumps in central regions, which may contribute to, but \added{may} not fully explain the observed central ``deadzone'' in clump age (Section~\ref{sec:radial}). All key trends in the main analysis were tested across a range of detection thresholds and remain consistent within the completeness limits (Section~\ref{sec:method}).

\added{
Because the clump mass function in Section \ref{sec:mass_function} is fit above $\log(M_\star/M_\odot)=8.2$, we examined whether residual incompleteness could bias the inferred slope. Although stellar mass was not directly assigned in the injection simulations, we estimated the effective detection probability of real clumps by combining their observed fluxes with the flux-dependent recovery fractions shown in Figure~\ref{fig:completeness}(d). For clumps with $\log(M_\star/M_\odot)\geq 8.2$, the mean detection probability is approximately $C_{\rm det}\sim0.6$--0.7 across redshift bins ($z\sim2.5$--5.5), with no strong systematic redshift dependence where statistics are robust. 

To test the stability of the inferred slope, we recomputed the mass function (i) restricting to $3<z<4$, where completeness is high, and (ii) adopting progressively higher mass thresholds of $\log(M_\star/M_\odot)=8.3$ and 8.4. The resulting slopes ($\alpha=-1.52$, $-1.54$, and $-1.47$, respectively) are consistent within uncertainties with our fiducial value of $\alpha=-1.50$. These tests indicate that residual incompleteness primarily affects the normalization of the mass function rather than its differential slope, and does not drive our physical interpretation of the clump mass function.
}

\section{Additional Galaxy Morphology Comparison}
\label{sec:app_gini_test}

\renewcommand{\thefigure}{B\arabic{figure}}
\setcounter{figure}{0}

\begin{figure*}[!ht]
    \centering
    \includegraphics[width=0.49\linewidth]{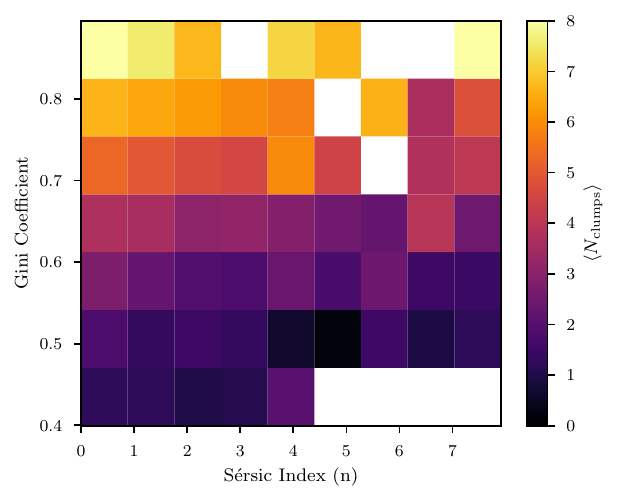}
    \includegraphics[width=0.49\linewidth]{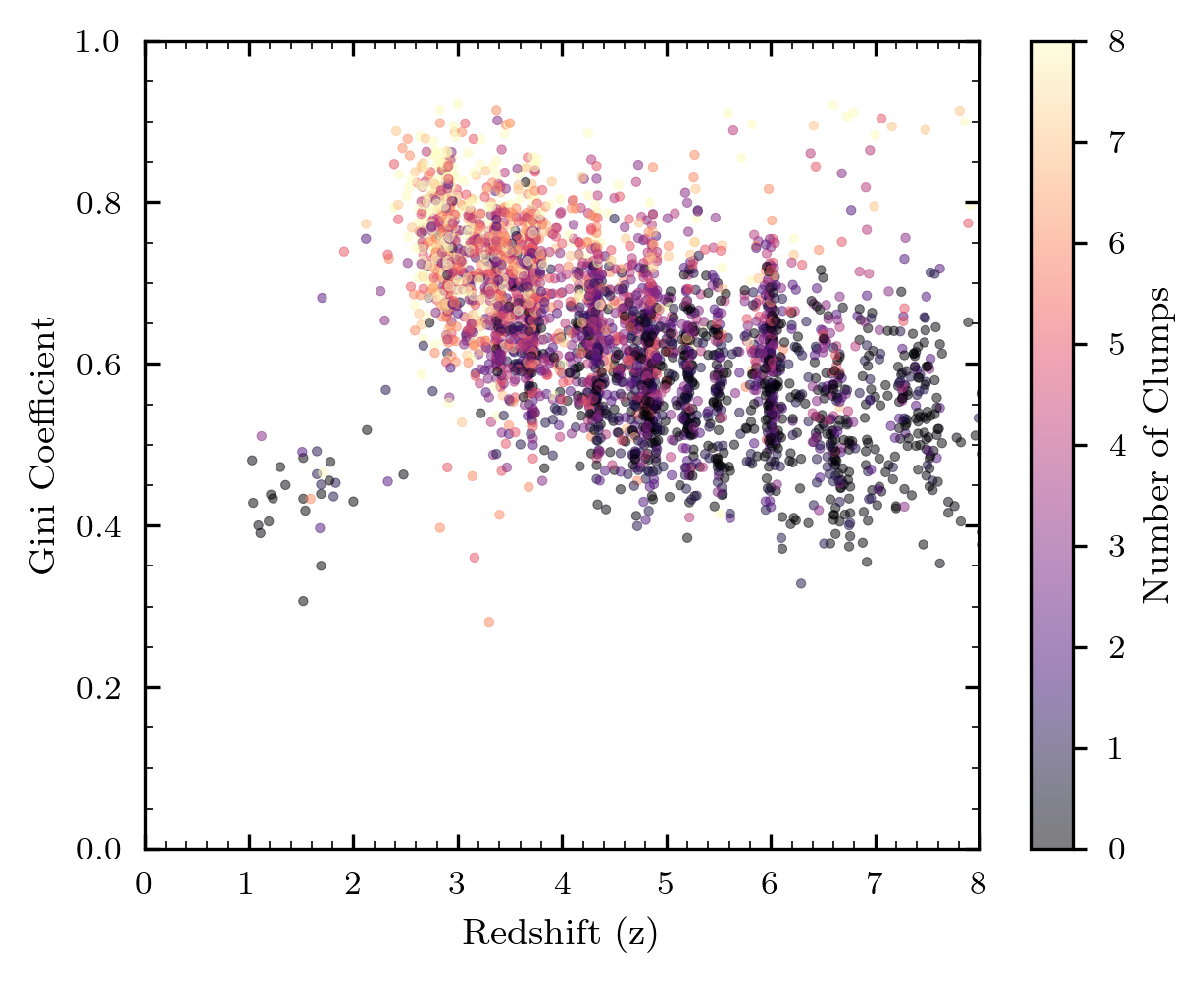}
    \caption{
Clump multiplicity as a function of global galaxy structural parameters and redshift. 
\textbf{Left:} Mean number of \added{clumps} per galaxy as a function of S\'ersic index ($n$) and Gini coefficient. 
Clumpy systems are not strongly dependent on S\'ersic index, but show a positive correlation with Gini, suggesting that the observed substructure may not be driven by central concentration but by multiple luminous regions distributed throughout the galaxy.
\textbf{Right:} Gini coefficient versus redshift, with points color-coded by the number of clumps. 
This reveals that the redshift evolution in clump frequency (seen in Figure~\ref{fig:clump_fraction}) may be partly driven by the underlying evolution in Gini, rather than redshift alone. 
These results support a physical interpretation where clumpiness arises from internal structural complexity, rather than simply from morphological compactness.
}
\label{fig:gini_test}
\end{figure*}

In the main text (Figure~\ref{fig:clump_fraction}), we showed that clump frequency increases toward $z \sim 2$ and correlates strongly with the Gini coefficient, while showing only a weak dependence on S\'ersic index. To further test whether this redshift trend is primarily driven by internal structure rather than cosmic time, we examine the joint dependence of clump multiplicity on Gini and S\'ersic $n$, as well as the redshift evolution of Gini itself.

Figure~\ref{fig:gini_test} presents two diagnostic plots. The left panel shows the mean number of clumps per galaxy as a function of both S\'ersic index and Gini coefficient. While clumpiness varies little with S\'ersic $n$ beyond $n \sim 2$, a strong positive trend with Gini is evident: galaxies with $G \gtrsim 0.7$ host 5--8 clumps on average, whereas those with $G \lesssim 0.5$ rarely host more than one. This supports the interpretation that high clump counts arise from internal structural complexity (e.g., multiple luminous knots) rather than a centrally concentrated profile.

To verify that this correlation is not a mathematical artifact of the Gini definition, we performed a simple test using mock galaxies composed of multiple identical Gaussian clumps with fixed total flux. The resulting Gini coefficients show no significant dependence on the number of clumps, indicating that the observed increase in Gini with clump multiplicity in the data is intrinsic rather than a consequence of pixel flux inequality. In other words, galaxies with higher Gini values genuinely exhibit more uneven and high-contrast substructure, rather than merely a greater number of components with comparable flux.

The right panel of Figure~\ref{fig:gini_test} shows Gini as a function of redshift, with individual galaxies color-coded by clump count. A steady increase in Gini from $z \sim 6$ to $z \sim 2$ closely tracks the rise in clump frequency, suggesting that the redshift evolution of clumpiness reflects structural changes in galaxies rather than cosmic time alone. These results reinforce the conclusion that clumpy morphologies are more strongly linked to internal light distribution (e.g., Gini) than to global shape parameters such as S\'ersic $n$. Studies of clump formation and evolution should therefore account for internal structural diversity, rather than attributing clump incidence solely to cosmic epoch or merger-driven assembly.

\section{Robustness to Host-Light Contamination}
\label{sec:app_host_contam}

\renewcommand{\thefigure}{C\arabic{figure}}
\setcounter{figure}{0}

\begin{figure*}[!ht]
\centering
\includegraphics[width=0.6\linewidth]{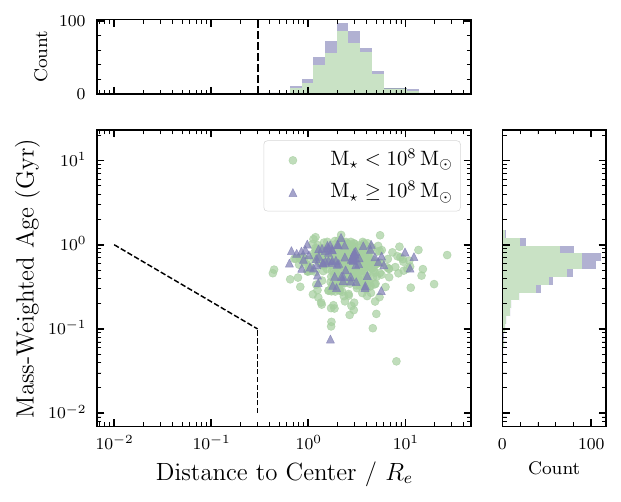}
\caption{
Same as Figure~\ref{fig:radial_profiles}, but restricted to clumps with high residual-to-total flux ratio, $f_{\rm res} > 0.5$.
}
\label{fig:radial_profiles_05}
\end{figure*}

\begin{figure*}[!ht]
\centering
\includegraphics[width=0.85\linewidth]{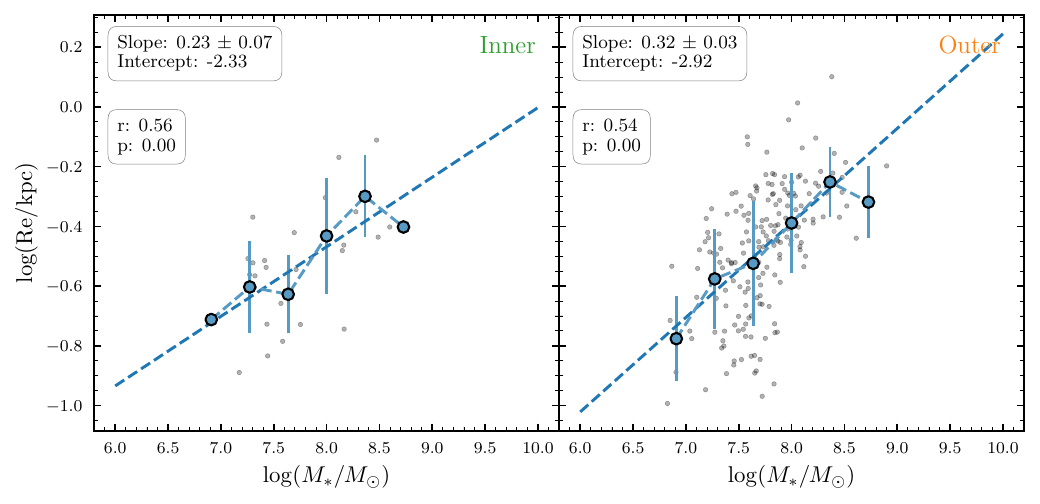}
\caption{
Same as Figure~\ref{fig:radial_smr}, but restricted to clumps with high residual-to-total flux ratio, $f_{\rm res} > 0.5$.
The slope difference between inner and outer clumps remains, with inner clumps showing a shallower size--mass relation than outer clumps.
}
\label{fig:radial_smr_05}
\end{figure*}

\added{
To assess the impact of host-galaxy light contamination on our clump property trends, we repeat the key radial analyses after restricting to clumps with high residual-to-total flux ratios. Specifically, for each segmented region we compute
\begin{equation}
f_{\rm res} \equiv \frac{F_{\rm res}}{F_{\rm orig}},
\end{equation}
where $F_{\rm res}$ is the total flux measured in the residual image and $F_{\rm orig}$ is the total flux in the original (pre-subtraction) image, both evaluated over the same segmented region (Section~\ref{sec:method}). We then require $f_{\rm res} > 0.5$, selecting clumps whose detected signal is dominated by flux in excess of the smooth S\'ersic model.

Figure~\ref{fig:radial_profiles_05} shows that the qualitative ``deadzone'' signature, a deficit of young clumps at small radii, remains visible under this conservative cut, supporting the interpretation that the central trend is not solely driven by host contamination. Figure~\ref{fig:radial_smr_05} shows that the inner versus outer size--mass relations are also consistent with the main results. In this restricted sample, the inferred slopes remain different, with inner clumps exhibiting a shallower relation than outer clumps, although with larger uncertainties due to reduced sample size. Overall, this test indicates that the principal radial trends reported in Section~\ref{sec:radial} persist when host contamination is minimized.}

\begin{thebibliography}{}
\expandafter\ifx\csname natexlab\endcsname\relax\def\natexlab#1{#1}\fi
\providecommand{\url}[1]{\href{#1}{#1}}
\providecommand{\dodoi}[1]{doi:~\href{http://doi.org/#1}{\nolinkurl{#1}}}
\providecommand{\doeprint}[1]{\href{http://ascl.net/#1}{\nolinkurl{http://ascl.net/#1}}}
\providecommand{\doarXiv}[1]{\href{https://arxiv.org/abs/#1}{\nolinkurl{https://arxiv.org/abs/#1}}}

\bibitem[{S. Alberts {et~al.}(2024)Alberts, Lyu, Shivaei, Rieke, Pérez-González, Bonaventura, Zhu, Helton, Ji, Morrison, Robertson, Stone, Sun, Williams, \& Willmer}]{alberts_smiles_2024}
Alberts, S., Lyu, J., Shivaei, I., {et~al.} 2024, \bibinfo{title}{{SMILES} {Initial} {Data} {Release}: {Unveiling} the {Obscured} {Universe} with {MIRI} {Multiband} {Imaging},} The Astrophysical Journal, 976, 224, \dodoi{10.3847/1538-4357/ad7396}

\bibitem[{N. Allen {et~al.}(2025)Allen, Oesch, Toft, Matharu, McPartland, Weibel, Brammer, Bowler, Ito, Gottumukkala, Rizzo, Valentino, Varadaraj, Weaver, \& Whitaker}]{allen_galaxy_2025}
Allen, N., Oesch, P.~A., Toft, S., {et~al.} 2025, \bibinfo{title}{Galaxy size and mass build-up in the first 2 {Gyr} of cosmic history from multi-wavelength {JWST} {NIRCam} imaging,} Astronomy and Astrophysics, 698, A30, \dodoi{10.1051/0004-6361/202452690}

\bibitem[{ {Astropy Collaboration} {et~al.}(2022){Astropy Collaboration}, Price-Whelan, Lim, Earl, Starkman, Bradley, Shupe, Patil, Corrales, Brasseur, Nöthe, Donath, Tollerud, Morris, Ginsburg, Vaher, Weaver, Tocknell, Jamieson, van Kerkwijk, Robitaille, Merry, Bachetti, Günther, Aldcroft, Alvarado-Montes, Archibald, Bódi, Bapat, Barentsen, Bazán, Biswas, Boquien, Burke, Cara, Cara, Conroy, Conseil, Craig, Cross, Cruz, D'Eugenio, Dencheva, Devillepoix, Dietrich, Eigenbrot, Erben, Ferreira, Foreman-Mackey, Fox, Freij, Garg, Geda, Glattly, Gondhalekar, Gordon, Grant, Greenfield, Groener, Guest, Gurovich, Handberg, Hart, Hatfield-Dodds, Homeier, Hosseinzadeh, Jenness, Jones, Joseph, Kalmbach, Karamehmetoglu, Kałuszyński, Kelley, Kern, Kerzendorf, Koch, Kulumani, Lee, Ly, Ma, MacBride, Maljaars, Muna, Murphy, Norman, O'Steen, Oman, Pacifici, Pascual, Pascual-Granado, Patil, Perren, Pickering, Rastogi, Roulston, Ryan, Rykoff, Sabater, Sakurikar, Salgado, Sanghi, Saunders, Savchenko, Schwardt,
  Seifert-Eckert, Shih, Jain, Shukla, Sick, Simpson, Singanamalla, Singer, Singhal, Sinha, Sipőcz, Spitler, Stansby, Streicher, Šumak, Swinbank, Taranu, Tewary, Tremblay, de~Val-Borro, Van~Kooten, Vasović, Verma, de~Miranda~Cardoso, Williams, Wilson, Winkel, Wood-Vasey, Xue, Yoachim, Zhang, Zonca, \& {Astropy Project Contributors}}]{astropy_collaboration_astropy_2022}
{Astropy Collaboration}, Price-Whelan, A.~M., Lim, P.~L., {et~al.} 2022, \bibinfo{title}{The {Astropy} {Project}: {Sustaining} and {Growing} a {Community}-oriented {Open}-source {Project} and the {Latest} {Major} {Release} (v5.0) of the {Core} {Package},} The Astrophysical Journal, 935, 167, \dodoi{10.3847/1538-4357/ac7c74}

\bibitem[{W.~M. Baker {et~al.}(2025)Baker, Tacchella, Johnson, Nelson, Suess, D'Eugenio, Curti, de~Graaff, Ji, Maiolino, Robertson, Scholtz, Alberts, Arribas, Boyett, Bunker, Carniani, Charlot, Chen, Chevallard, Curtis-Lake, Danhaive, DeCoursey, Egami, Eisenstein, Endsley, Hausen, Helton, Kumari, Looser, Maseda, Puskás, Rieke, Sandles, Sun, Übler, Williams, Willmer, \& Witstok}]{baker_core_2025}
Baker, W.~M., Tacchella, S., Johnson, B.~D., {et~al.} 2025, \bibinfo{title}{A core in a star-forming disc as evidence of inside-out growth in the early {Universe},} Nature Astronomy, 9, 141, \dodoi{10.1038/s41550-024-02384-8}

\bibitem[{F. Bournaud {et~al.}(2007)Bournaud, Elmegreen, \& Elmegreen}]{bournaud_rapid_2007}
Bournaud, F., Elmegreen, B.~G., \& Elmegreen, D.~M. 2007, \bibinfo{title}{Rapid {Formation} of {Exponential} {Disks} and {Bulges} at {High} {Redshift} from the {Dynamical} {Evolution} of {Clump}-{Cluster} and {Chain} {Galaxies},} The Astrophysical Journal, 670, 237, \dodoi{10.1086/522077}

\bibitem[{F. Bournaud {et~al.}(2014)Bournaud, Perret, Renaud, Dekel, Elmegreen, Elmegreen, Teyssier, Amram, Daddi, Duc, Elbaz, Epinat, Gabor, Juneau, Kraljic, \& Le~Floch'}]{bournaud_long_2014}
Bournaud, F., Perret, V., Renaud, F., {et~al.} 2014, \bibinfo{title}{The {Long} {Lives} of {Giant} {Clumps} and the {Birth} of {Outflows} in {Gas}-rich {Galaxies} at {High} {Redshift},} The Astrophysical Journal, 780, 57, \dodoi{10.1088/0004-637X/780/1/57}

\bibitem[{G.~B. Brammer {et~al.}(2008)Brammer, van Dokkum, \& Coppi}]{brammer_eazy_2008}
Brammer, G.~B., van Dokkum, P.~G., \& Coppi, P. 2008, \bibinfo{title}{{EAZY}: {A} {Fast}, {Public} {Photometric} {Redshift} {Code},} The Astrophysical Journal, 686, 1503, \dodoi{10.1086/591786}

\bibitem[{A.~J. Bunker {et~al.}(2024)Bunker, Cameron, Curtis-Lake, Jakobsen, Carniani, Curti, Witstok, Maiolino, D'Eugenio, Looser, Willott, Bonaventura, Hainline, Übler, Willmer, Saxena, Smit, Alberts, Arribas, Baker, Baum, Bhatawdekar, Bowler, Boyett, Charlot, Chen, Chevallard, Circosta, DeCoursey, de~Graaff, Egami, Eisenstein, Endsley, Ferruit, Giardino, Hausen, Helton, Hviding, Ji, Johnson, Jones, Kumari, Laseter, Lützgendorf, Maseda, Nelson, Parlanti, Perna, Rauscher, Rawle, Rix, Rieke, Robertson, Rodríguez Del~Pino, Sandles, Scholtz, Sharpe, Skarbinski, Stark, Sun, Tacchella, Topping, Villanueva, Wallace, Williams, \& Woodrum}]{bunker_jades_2024}
Bunker, A.~J., Cameron, A.~J., Curtis-Lake, E., {et~al.} 2024, \bibinfo{title}{{JADES} {NIRSpec} initial data release for the {Hubble} {Ultra} {Deep} {Field}: {Redshifts} and line fluxes of distant galaxies from the deepest {JWST} {Cycle} 1 {NIRSpec} multi-object spectroscopy,} Astronomy and Astrophysics, 690, A288, \dodoi{10.1051/0004-6361/202347094}

\bibitem[{A. Calabrò {et~al.}(2019)Calabrò, Daddi, Fensch, Bournaud, Cibinel, Puglisi, Jin, Delvecchio, \& D'Eugenio}]{calabro_merger_2019}
Calabrò, A., Daddi, E., Fensch, J., {et~al.} 2019, \bibinfo{title}{Merger induced clump formation in distant infrared luminous starburst galaxies,} Astronomy and Astrophysics, 632, A98, \dodoi{10.1051/0004-6361/201935778}

\bibitem[{D. Ceverino {et~al.}(2010)Ceverino, Dekel, \& Bournaud}]{ceverino_high-redshift_2010}
Ceverino, D., Dekel, A., \& Bournaud, F. 2010, \bibinfo{title}{High-redshift clumpy discs and bulges in cosmological simulations,} Monthly Notices of the Royal Astronomical Society, 404, 2151, \dodoi{10.1111/j.1365-2966.2010.16433.x}

\bibitem[{D. Ceverino {et~al.}(2012)Ceverino, Dekel, Mandelker, Bournaud, Burkert, Genzel, \& Primack}]{ceverino_rotational_2012}
Ceverino, D., Dekel, A., Mandelker, N., {et~al.} 2012, \bibinfo{title}{Rotational support of giant clumps in high-z disc galaxies,} Monthly Notices of the Royal Astronomical Society, 420, 3490, \dodoi{10.1111/j.1365-2966.2011.20296.x}

\bibitem[{G. Chabrier(2003)Chabrier}]{chabrier_galactic_2003}
Chabrier, G. 2003, \bibinfo{title}{Galactic {Stellar} and {Substellar} {Initial} {Mass} {Function},} Publications of the Astronomical Society of the Pacific, 115, 763, \dodoi{10.1086/376392}

\bibitem[{A. Claeyssens {et~al.}(2023)Claeyssens, Adamo, Richard, Mahler, Messa, \& Dessauges-Zavadsky}]{claeyssens_star_2023}
Claeyssens, A., Adamo, A., Richard, J., {et~al.} 2023, \bibinfo{title}{Star formation at the smallest scales: a {JWST} study of the clump populations in {SMACS0723},} Monthly Notices of the Royal Astronomical Society, 520, 2180, \dodoi{10.1093/mnras/stac3791}

\bibitem[{E. Curtis-Lake {et~al.}(2025)Curtis-Lake, Cameron, Bunker, Scholtz, Carniani, Parlanti, D'Eugenio, Jakobsen, Willmer, Arribas, Baker, Charlot, Chevallard, Circosta, Curti, Eisenstein, Hainline, Ji, Johnson, Jones, Maiolino, Maseda, Pérez-González, Rawle, Rieke, Rinaldi, Robertson, Rodrígez Del~Pino, Saxena, Shivaei, Smit, Tacchella, Übler, Venturi, Williams, Willott, \& Duan}]{curtis-lake_jades_2025}
Curtis-Lake, E., Cameron, A.~J., Bunker, A.~J., {et~al.} 2025, \bibinfo{title}{{JADES} {Data} {Release} 4 {Paper} {I}: {Sample} {Selection}, {Observing} {Strategy} and {Redshifts} of the complete spectroscopic sample,} arXiv, \dodoi{10.48550/arXiv.2510.01033}

\bibitem[{A.~L. Danhaive {et~al.}(2025{\natexlab{a}})Danhaive, Tacchella, Übler, de~Graaff, Egami, Johnson, Sun, Arribas, Bunker, Carniani, Jones, Maiolino, McClymont, Parlanti, Simmonds, Villanueva, Baker, Jaffe, Eisenstein, Hainline, Helton, Ji, Lin, Liu, Puskás, Rieke, Rinaldi, Robertson, Scholz, Williams, \& Willmer}]{danhaive_dawn_2025}
Danhaive, A.~L., Tacchella, S., Übler, H., {et~al.} 2025{\natexlab{a}}, \bibinfo{title}{The dawn of discs: unveiling the turbulent ionized gas kinematics of the galaxy population at z ∼ 4─6 with {JWST}/{NIRCam} grism spectroscopy,} Monthly Notices of the Royal Astronomical Society, 543, 3249, \dodoi{10.1093/mnras/staf1540}

\bibitem[{A.~L. Danhaive {et~al.}(2025{\natexlab{b}})Danhaive, Tacchella, McClymont, Robertson, Carniani, Carreira, Egami, Bunker, Curtis-Lake, Eisenstein, Ji, Johnson, Rieke, Villanueva, Willmer, Willot, Wu, \& Zhu}]{danhaive_beyond_2025}
Danhaive, A.~L., Tacchella, S., McClymont, W., {et~al.} 2025{\natexlab{b}}, \bibinfo{title}{Beyond the stars: {Linking} {H}\$α\$ sizes, kinematics, and star formation in galaxies at \$z{\textbackslash}approx 4-6\$ with {JWST} grism surveys and \${\textbackslash}texttt\{geko\}\$,} arXiv, \dodoi{10.48550/arXiv.2510.06315}

\bibitem[{A. Dekel {et~al.}(2022)Dekel, Mandelker, Bournaud, Ceverino, Guo, \& Primack}]{dekel_clump_2022}
Dekel, A., Mandelker, N., Bournaud, F., {et~al.} 2022, \bibinfo{title}{Clump survival and migration in {VDI} galaxies: an analytical model versus simulations and observations,} Monthly Notices of the Royal Astronomical Society, 511, 316, \dodoi{10.1093/mnras/stab3810}

\bibitem[{A. Dekel {et~al.}(2009)Dekel, Birnboim, Engel, Freundlich, Goerdt, Mumcuoglu, Neistein, Pichon, Teyssier, \& Zinger}]{dekel_cold_2009}
Dekel, A., Birnboim, Y., Engel, G., {et~al.} 2009, \bibinfo{title}{Cold streams in early massive hot haloes as the main mode of galaxy formation,} Nature, 457, 451, \dodoi{10.1038/nature07648}

\bibitem[{M. Dessauges-Zavadsky \& A. Adamo(2018)Dessauges-Zavadsky \& Adamo}]{dessauges-zavadsky_first_2018}
Dessauges-Zavadsky, M., \& Adamo, A. 2018, \bibinfo{title}{First constraints on the stellar mass function of star-forming clumps at the peak of cosmic star formation,} Monthly Notices of the Royal Astronomical Society, 479, L118, \dodoi{10.1093/mnrasl/sly112}

\bibitem[{F. D'Eugenio {et~al.}(2025)D'Eugenio, Cameron, Scholtz, Carniani, Willott, Curtis-Lake, Bunker, Parlanti, Maiolino, Willmer, Jakobsen, Robertson, Johnson, Tacchella, Cargile, Rawle, Arribas, Chevallard, Curti, Egami, Eisenstein, Kumari, Looser, Rieke, Rodríguez Del~Pino, Saxena, Übler, Venturi, Witstok, Baker, Bhatawdekar, Bonaventura, Boyett, Charlot, Danhaive, Hainline, Hausen, Helton, Ji, Ji, Jones, Juodžbalis, Maseda, Pérez-González, Perna, Puskás, Shivaei, Silcock, Simmonds, Smit, Sun, Villanueva, Williams, \& Zhu}]{deugenio_jades_2025}
D'Eugenio, F., Cameron, A.~J., Scholtz, J., {et~al.} 2025, \bibinfo{title}{{JADES} {Data} {Release} 3: {NIRSpec}/{Microshutter} {Assembly} {Spectroscopy} for 4000 {Galaxies} in the {GOODS} {Fields},} The Astrophysical Journal Supplement Series, 277, 4, \dodoi{10.3847/1538-4365/ada148}

\bibitem[{Q. Duan {et~al.}(2025)Duan, Conselice, Li, Austin, Harvey, Adams, Duncan, Trussler, Ferreira, Westcott, Harris, Windhorst, Holwerda, Broadhurst, Coe, Cohen, Du, Driver, Frye, Grogin, Hathi, Jansen, Koekemoer, Marshall, Nonino, Ortiz, Pirzkal, Robotham, Ryan, Summers, D'Silva, Willmer, \& Yan}]{duan_galaxy_2025}
Duan, Q., Conselice, C.~J., Li, Q., {et~al.} 2025, \bibinfo{title}{Galaxy mergers in the epoch of reionization – {I}. {A} {JWST} study of pair fractions, merger rates, and stellar mass accretion rates at z = 4.5–11.5,} Monthly Notices of the Royal Astronomical Society, 540, 774, \dodoi{10.1093/mnras/staf638}

\bibitem[{D.~J. Eisenstein {et~al.}(2023)Eisenstein, Willott, Alberts, Arribas, Bonaventura, Bunker, Cameron, Carniani, Charlot, Curtis-Lake, D'Eugenio, Endsley, Ferruit, Giardino, Hainline, Hausen, Jakobsen, Johnson, Maiolino, Rieke, Rieke, Rix, Robertson, Stark, Tacchella, Williams, Willmer, Baker, Baum, Bhatawdekar, Boyett, Chen, Chevallard, Circosta, Curti, Danhaive, DeCoursey, de~Graaff, Dressler, Egami, Helton, Hviding, Ji, Jones, Kumari, Lützgendorf, Laseter, Looser, Lyu, Maseda, Nelson, Parlanti, Perna, Puskás, Rawle, Del~Pino, Sandles, Saxena, Scholtz, Sharpe, Shivaei, Silcock, Simmonds, Skarbinski, Smit, Stone, Suess, Sun, Tang, Topping, Übler, Villanueva, Wallace, Whitler, Witstok, \& Woodrum}]{eisenstein_overview_2023}
Eisenstein, D.~J., Willott, C., Alberts, S., {et~al.} 2023, \bibinfo{title}{Overview of the {JWST} {Advanced} {Deep} {Extragalactic} {Survey} ({JADES}),} arXiv.
\newblock \url{http://arxiv.org/abs/2306.02465}

\bibitem[{D.~J. Eisenstein {et~al.}(2025)Eisenstein, Johnson, Robertson, Tacchella, Hainline, Jakobsen, Maiolino, Bonaventura, Bunker, Cameron, Cargile, Curtis-Lake, Hausen, Puskás, Rieke, Sun, Willmer, Willott, Alberts, Arribas, Baker, Baum, Bhatawdekar, Carniani, Charlot, Chen, Chevallard, Curti, DeCoursey, D'Eugenio, de~Graaff, Egami, Helton, Ji, Jones, Kumari, Lützgendorf, Laseter, Looser, Lyu, Maseda, Nelson, Parlanti, Rauscher, Rawle, Rieke, Rix, Rujopakarn, Sandles, Saxena, Scholtz, Sharpe, Shivaei, Simmonds, Smit, Topping, Übler, Venturi, Williams, Witstok, \& Woodrum}]{eisenstein_jades_2025}
Eisenstein, D.~J., Johnson, B.~D., Robertson, B., {et~al.} 2025, \bibinfo{title}{The {JADES} {Origins} {Field}: {A} {New} {JWST} {Deep} {Field} in the {JADES} {Second} {NIRCam} {Data} {Release},} The Astrophysical Journal Supplement Series, 281, 50, \dodoi{10.3847/1538-4365/ae1137}

\bibitem[{B.~G. Elmegreen {et~al.}(2006)Elmegreen, Elmegreen, Chandar, Whitmore, \& Regan}]{elmegreen_hierarchical_2006}
Elmegreen, B.~G., Elmegreen, D.~M., Chandar, R., Whitmore, B., \& Regan, M. 2006, \bibinfo{title}{Hierarchical {Star} {Formation} in the {Spiral} {Galaxy} {NGC} 628,} The Astrophysical Journal, 644, 879, \dodoi{10.1086/503797}

\bibitem[{D.~M. Elmegreen {et~al.}(2009)Elmegreen, Elmegreen, Marcus, Shahinyan, Yau, \& Petersen}]{elmegreen_clumpy_2009}
Elmegreen, D.~M., Elmegreen, B.~G., Marcus, M.~T., {et~al.} 2009, \bibinfo{title}{Clumpy {Galaxies} in {Goods} and {Gems}: {Massive} {Analogs} of {Local} {Dwarf} {Irregulars},} The Astrophysical Journal, 701, 306, \dodoi{10.1088/0004-637X/701/1/306}

\bibitem[{D.~M. Elmegreen {et~al.}(2004)Elmegreen, Elmegreen, \& Sheets}]{elmegreen_chain_2004}
Elmegreen, D.~M., Elmegreen, B.~G., \& Sheets, C.~M. 2004, \bibinfo{title}{Chain {Galaxies} in the {Tadpole} {Advanced} {Camera} for {Surveys} {Field},} The Astrophysical Journal, 603, 74, \dodoi{10.1086/381357}

\bibitem[{N.~M. Förster~Schreiber {et~al.}(2011)Förster~Schreiber, Shapley, Genzel, Bouché, Cresci, Davies, Erb, Genel, Lutz, Newman, Shapiro, Steidel, Sternberg, \& Tacconi}]{forster_schreiber_constraints_2011}
Förster~Schreiber, N.~M., Shapley, A.~E., Genzel, R., {et~al.} 2011, \bibinfo{title}{Constraints on the {Assembly} and {Dynamics} of {Galaxies}. {II}. {Properties} of {Kiloparsec}-scale {Clumps} in {Rest}-frame {Optical} {Emission} of z {\textasciitilde} 2 {Star}-forming {Galaxies},} The Astrophysical Journal, 739, 45, \dodoi{10.1088/0004-637X/739/1/45}

\bibitem[{S. Genel {et~al.}(2012)Genel, Naab, Genzel, Förster~Schreiber, Sternberg, Oser, Johansson, Davé, Oppenheimer, \& Burkert}]{genel_short-lived_2012}
Genel, S., Naab, T., Genzel, R., {et~al.} 2012, \bibinfo{title}{Short-lived {Star}-forming {Giant} {Clumps} in {Cosmological} {Simulations} of z ≈ 2 {Disks},} The Astrophysical Journal, 745, 11, \dodoi{10.1088/0004-637X/745/1/11}

\bibitem[{R. Genzel {et~al.}(2011)Genzel, Newman, Jones, Förster~Schreiber, Shapiro, Genel, Lilly, Renzini, Tacconi, Bouché, Burkert, Cresci, Buschkamp, Carollo, Ceverino, Davies, Dekel, Eisenhauer, Hicks, Kurk, Lutz, Mancini, Naab, Peng, Sternberg, Vergani, \& Zamorani}]{genzel_sins_2011}
Genzel, R., Newman, S., Jones, T., {et~al.} 2011, \bibinfo{title}{The {Sins} {Survey} of z {\textasciitilde} 2 {Galaxy} {Kinematics}: {Properties} of the {Giant} {Star}-forming {Clumps},} The Astrophysical Journal, 733, 101, \dodoi{10.1088/0004-637X/733/2/101}

\bibitem[{Y. Guo {et~al.}(2012)Guo, Giavalisco, Ferguson, Cassata, \& Koekemoer}]{guo_multi-wavelength_2012}
Guo, Y., Giavalisco, M., Ferguson, H.~C., Cassata, P., \& Koekemoer, A.~M. 2012, \bibinfo{title}{Multi-wavelength {View} of {Kiloparsec}-scale {Clumps} in {Star}-forming {Galaxies} at z {\textasciitilde} 2,} The Astrophysical Journal, 757, 120, \dodoi{10.1088/0004-637X/757/2/120}

\bibitem[{Y. Guo {et~al.}(2015)Guo, Ferguson, Bell, Koo, Conselice, Giavalisco, Kassin, Lu, Lucas, Mandelker, McIntosh, Primack, Ravindranath, Barro, Ceverino, Dekel, Faber, Fang, Koekemoer, Noeske, Rafelski, \& Straughn}]{guo_clumpy_2015}
Guo, Y., Ferguson, H.~C., Bell, E.~F., {et~al.} 2015, \bibinfo{title}{Clumpy {Galaxies} in {CANDELS}. {I}. {The} {Definition} of {UV} {Clumps} and the {Fraction} of {Clumpy} {Galaxies} at 0.5 {\textless} z {\textless} 3,} The Astrophysical Journal, 800, 39, \dodoi{10.1088/0004-637X/800/1/39}

\bibitem[{K.~N. Hainline {et~al.}(2024)Hainline, Johnson, Robertson, Tacchella, Helton, Sun, Eisenstein, Simmonds, Topping, Whitler, Willmer, Rieke, Suess, Hviding, Cameron, Alberts, Baker, Baum, Bhatawdekar, Bonaventura, Boyett, Bunker, Carniani, Charlot, Chevallard, Chen, Curti, Curtis-Lake, D'Eugenio, Egami, Endsley, Hausen, Ji, Looser, Lyu, Maiolino, Nelson, Puskás, Rawle, Sandles, Saxena, Smit, Stark, Williams, Willott, \& Witstok}]{hainline_cosmos_2024}
Hainline, K.~N., Johnson, B.~D., Robertson, B., {et~al.} 2024, \bibinfo{title}{The {Cosmos} in {Its} {Infancy}: {JADES} {Galaxy} {Candidates} at z {\textgreater} 8 in {GOODS}-{S} and {GOODS}-{N},} The Astrophysical Journal, 964, 71, \dodoi{10.3847/1538-4357/ad1ee4}

\bibitem[{J.~M. Helton {et~al.}(2024)Helton, Sun, Woodrum, Hainline, Willmer, Rieke, Rieke, Alberts, Eisenstein, Tacchella, Robertson, Johnson, Baker, Bhatawdekar, Bunker, Chen, Egami, Ji, Maiolino, Willott, \& Witstok}]{helton_identification_2024}
Helton, J.~M., Sun, F., Woodrum, C., {et~al.} 2024, \bibinfo{title}{Identification of {High}-redshift {Galaxy} {Overdensities} in {GOODS}-{N} and {GOODS}-{S},} The Astrophysical Journal, 974, 41, \dodoi{10.3847/1538-4357/ad6867}

\bibitem[{P.~F. Hopkins(2013)Hopkins}]{hopkins_general_2013}
Hopkins, P.~F. 2013, \bibinfo{title}{A general theory of turbulent fragmentation,} Monthly Notices of the Royal Astronomical Society, 430, 1653, \dodoi{10.1093/mnras/sts704}

\bibitem[{J.~D. Hunter(2007)Hunter}]{hunter_matplotlib_2007}
Hunter, J.~D. 2007, \bibinfo{title}{Matplotlib: {A} {2D} {Graphics} {Environment},} Computing in Science and Engineering, 9, 90, \dodoi{10.1109/MCSE.2007.55}

\bibitem[{R. Ikeda {et~al.}(2025)Ikeda, Iono, Tadaki, Franco, Yun, Zavala, Tamura, Tsukui, Williams, Hatsukade, Lee, Michiyama, Mitsuhashi, Nakanishi, Casey, Ikarashi, Lee, Matsuda, Saito, Silva, Umehata, \& Yajima}]{ikeda_formation_2025}
Ikeda, R., Iono, D., Tadaki, K.-i., {et~al.} 2025, \bibinfo{title}{Formation {Of} {Sub}-{Structure} {In} {Luminous} {Submillimeter} galaxies ({FOSSILS}): {Evidence} of {Multiple} {Pathways} to {Trigger} {Starbursts} in {Luminous} {Submillimeter} {Galaxies},} arXiv, \dodoi{10.48550/arXiv.2510.18006}

\bibitem[{A. Immeli {et~al.}(2004)Immeli, Samland, Gerhard, \& Westera}]{immeli_gas_2004}
Immeli, A., Samland, M., Gerhard, O., \& Westera, P. 2004, \bibinfo{title}{Gas physics, disk fragmentation, and bulge formation in young galaxies,} Astronomy and Astrophysics, 413, 547, \dodoi{10.1051/0004-6361:20034282}

\bibitem[{S. Inoue {et~al.}(2016)Inoue, Dekel, Mandelker, Ceverino, Bournaud, \& Primack}]{inoue_non-linear_2016}
Inoue, S., Dekel, A., Mandelker, N., {et~al.} 2016, \bibinfo{title}{Non-linear violent disc instability with high {Toomre}'s {Q} in high-redshift clumpy disc galaxies,} Monthly Notices of the Royal Astronomical Society, 456, 2052, \dodoi{10.1093/mnras/stv2793}

\bibitem[{S. Inoue \& T.~R. Saitoh(2012)Inoue \& Saitoh}]{inoue_natures_2012}
Inoue, S., \& Saitoh, T.~R. 2012, \bibinfo{title}{Natures of a clump-origin bulge: a pseudo-bulge like but old metal-rich bulge,} Monthly Notices of the Royal Astronomical Society, 422, 1902, \dodoi{10.1111/j.1365-2966.2011.20338.x}

\bibitem[{B.~D. Johnson {et~al.}(2021)Johnson, Leja, Conroy, \& Speagle}]{johnson_stellar_2021}
Johnson, B.~D., Leja, J., Conroy, C., \& Speagle, J.~S. 2021, \bibinfo{title}{Stellar {Population} {Inference} with {Prospector},} The Astrophysical Journal Supplement Series, 254, 22, \dodoi{10.3847/1538-4365/abef67}

\bibitem[{B.~S. Kalita {et~al.}(2025{\natexlab{a}})Kalita, Silverman, Daddi, Mercier, Ho, \& Ding}]{kalita_near-ir_2025}
Kalita, B.~S., Silverman, J.~D., Daddi, E., {et~al.} 2025{\natexlab{a}}, \bibinfo{title}{Near-{IR} clumps and their properties in high-z galaxies with {JWST}/{NIRCam},} Monthly Notices of the Royal Astronomical Society, 537, 402, \dodoi{10.1093/mnras/staf031}

\bibitem[{B.~S. Kalita {et~al.}(2025{\natexlab{b}})Kalita, Suzuki, Kashino, Silverman, Daddi, Ho, Ding, Mercier, Faisst, Sheth, Valentino, Puglisi, Saito, Kakkad, Ilbert, Khostovan, Liu, Tanaka, Magdis, Zavala, Tan, Kartaltepe, Yang, Koekemoer, McKinney, Robertson, Jin, Hayward, Hirschmann, Franco, Shuntov, Gozaliasl, Kaminsky, \& Rich}]{kalita_clumps_2025}
Kalita, B.~S., Suzuki, T.~L., Kashino, D., {et~al.} 2025{\natexlab{b}}, \bibinfo{title}{Clumps as multiscale structures in cosmic noon galaxies,} Monthly Notices of the Royal Astronomical Society, 536, 3090, \dodoi{10.1093/mnras/stae2781}

\bibitem[{M.~R. Krumholz {et~al.}(2018)Krumholz, Burkhart, Forbes, \& Crocker}]{krumholz_unified_2018}
Krumholz, M.~R., Burkhart, B., Forbes, J.~C., \& Crocker, R.~M. 2018, \bibinfo{title}{A unified model for galactic discs: star formation, turbulence driving, and mass transport,} Monthly Notices of the Royal Astronomical Society, 477, 2716, \dodoi{10.1093/mnras/sty852}

\bibitem[{J. Leja {et~al.}(2019)Leja, Carnall, Johnson, Conroy, \& Speagle}]{leja_how_2019}
Leja, J., Carnall, A.~C., Johnson, B.~D., Conroy, C., \& Speagle, J.~S. 2019, \bibinfo{title}{How to {Measure} {Galaxy} {Star} {Formation} {Histories}. {II}. {Nonparametric} {Models},} The Astrophysical Journal, 876, 3, \dodoi{10.3847/1538-4357/ab133c}

\bibitem[{L. Liu {et~al.}(2022)Liu, Bureau, Li, Davis, Nguyen, Liang, Choi, Smith, \& Iguchi}]{liu_wisdom_2022}
Liu, L., Bureau, M., Li, G.-X., {et~al.} 2022, \bibinfo{title}{{WISDOM} {Project} - {XII}. {Clump} properties and turbulence regulated by clump-clump collisions in the dwarf galaxy {NGC} 404,} Monthly Notices of the Royal Astronomical Society, 517, 632, \dodoi{10.1093/mnras/stac2287}

\bibitem[{R.~C. Livermore {et~al.}(2015)Livermore, Jones, Richard, Bower, Swinbank, Yuan, Edge, Ellis, Kewley, Smail, Coppin, \& Ebeling}]{livermore_resolved_2015}
Livermore, R.~C., Jones, T.~A., Richard, J., {et~al.} 2015, \bibinfo{title}{Resolved spectroscopy of gravitationally lensed galaxies: global dynamics and star-forming clumps on ∼100 pc scales at 1 {\textless} z {\textless} 4,} Monthly Notices of the Royal Astronomical Society, 450, 1812, \dodoi{10.1093/mnras/stv686}

\bibitem[{J.~M. Lotz {et~al.}(2004)Lotz, Primack, \& Madau}]{lotz_new_2004}
Lotz, J.~M., Primack, J., \& Madau, P. 2004, \bibinfo{title}{A {New} {Nonparametric} {Approach} to {Galaxy} {Morphological} {Classification},} The Astronomical Journal, 128, 163, \dodoi{10.1086/421849}

\bibitem[{N. Mandelker {et~al.}(2017)Mandelker, Dekel, Ceverino, DeGraf, Guo, \& Primack}]{mandelker_giant_2017}
Mandelker, N., Dekel, A., Ceverino, D., {et~al.} 2017, \bibinfo{title}{Giant clumps in simulated high- z {Galaxies}: properties, evolution and dependence on feedback,} Monthly Notices of the Royal Astronomical Society, 464, 635, \dodoi{10.1093/mnras/stw2358}

\bibitem[{N. Mandelker {et~al.}(2014)Mandelker, Dekel, Ceverino, Tweed, Moody, \& Primack}]{mandelker_population_2014}
Mandelker, N., Dekel, A., Ceverino, D., {et~al.} 2014, \bibinfo{title}{The population of giant clumps in simulated high-z galaxies: in situ and ex situ migration and survival,} Monthly Notices of the Royal Astronomical Society, 443, 3675, \dodoi{10.1093/mnras/stu1340}

\bibitem[{E.~P. Mathews {et~al.}(2023)Mathews, Leja, Speagle, Johnson, Gibson, Nelson, Suess, Tacchella, Whitaker, \& Wang}]{mathews_as_2023}
Mathews, E.~P., Leja, J., Speagle, J.~S., {et~al.} 2023, \bibinfo{title}{As {Simple} as {Possible} but {No} {Simpler}: {Optimizing} the {Performance} of {Neural} {Net} {Emulators} for {Galaxy} {SED} {Fitting},} The Astrophysical Journal, 954, 132, \dodoi{10.3847/1538-4357/ace720}

\bibitem[{K. Mawatari {et~al.}(2025)Mawatari, Costantin, Usui, Hashimoto, Álvarez Márquez, Sugahara, Colina, Inoue, Osone, Arribas, Marques-Chaves, Nakazato, Hagimoto, Hashigaya, Ceverino, Yoshida, Bakx, Fudamoto, Crespo~Gómez, Matsuo, Pereira-Santaella, Blanco-Prieto, Ren, \& Tamura}]{mawatari_rioja_2025}
Mawatari, K., Costantin, L., Usui, M., {et~al.} 2025, \bibinfo{title}{{RIOJA}. {Merger}-induced {Clumps} in a {Galaxy} at {Redshift} 6.81 {Revealed} by {JWST},} arXiv, \dodoi{10.48550/arXiv.2507.02053}

\bibitem[{F. Meyer(1994)Meyer}]{meyer_topographic_1994}
Meyer, F. 1994, \bibinfo{title}{Topographic distance and watershed lines,} Signal Processing, 38, 113, \dodoi{10.1016/0165-1684(94)90060-4}

\bibitem[{T.~B. Miller {et~al.}(2025)Miller, Suess, Setton, Price, Labbe, Bezanson, Brammer, Cutler, Furtak, Leja, Pan, Wang, Weaver, Whitaker, Dayal, de~Graaff, Feldmann, Greene, Fujimoto, Maseda, Nanayakkara, Nelson, van Dokkum, \& Zitrin}]{miller_jwst_2025}
Miller, T.~B., Suess, K.~A., Setton, D.~J., {et~al.} 2025, \bibinfo{title}{{JWST} {UNCOVERs} the {Optical} {Size}–{Stellar} {Mass} {Relation} at 4 {\textless} z {\textless} 8: {Rapid} {Growth} in the {Sizes} of {Low}-mass {Galaxies} in the {First} {Billion} {Years} of the {Universe},} The Astrophysical Journal, 988, 196, \dodoi{10.3847/1538-4357/ade438}

\bibitem[{Y. Nakazato {et~al.}(2024)Nakazato, Ceverino, \& Yoshida}]{nakazato_merger-driven_2024}
Nakazato, Y., Ceverino, D., \& Yoshida, N. 2024, \bibinfo{title}{A {Merger}-driven {Scenario} for {Clumpy} {Galaxy} {Formation} in the {Epoch} of {Reionization}: {Physical} {Properties} of {Clumps} in the {FirstLight} {Simulation},} The Astrophysical Journal, 975, 238, \dodoi{10.3847/1538-4357/ad7d0b}

\bibitem[{S.~F. Newman {et~al.}(2012)Newman, Shapiro~Griffin, Genzel, Davies, Förster-Schreiber, Tacconi, Kurk, Wuyts, Genel, Lilly, Renzini, Bouché, Burkert, Cresci, Buschkamp, Carollo, Eisenhauer, Hicks, Lutz, Mancini, Naab, Peng, \& Vergani}]{newman_shocked_2012}
Newman, S.~F., Shapiro~Griffin, K., Genzel, R., {et~al.} 2012, \bibinfo{title}{Shocked {Superwinds} from the z {\textasciitilde} 2 {Clumpy} {Star}-forming {Galaxy}, {ZC406690},} The Astrophysical Journal, 752, 111, \dodoi{10.1088/0004-637X/752/2/111}

\bibitem[{P.~A. Oesch {et~al.}(2023)Oesch, Brammer, Naidu, Bouwens, Chisholm, Illingworth, Matthee, Nelson, Qin, Reddy, Shapley, Shivaei, van Dokkum, Weibel, Whitaker, Wuyts, Covelo-Paz, Endsley, Fudamoto, Giovinazzo, Herard-Demanche, Kerutt, Kramarenko, Labbe, Leonova, Lin, Magee, Marchesini, Maseda, Mason, Matharu, Meyer, Neufeld, Prieto~Lyon, Schaerer, Sharma, Shuntov, Smit, Stefanon, Wyithe, \& Xiao}]{oesch_jwst_2023}
Oesch, P.~A., Brammer, G., Naidu, R.~P., {et~al.} 2023, \bibinfo{title}{The {JWST} {FRESCO} survey: legacy {NIRCam}/grism spectroscopy and imaging in the two {GOODS} fields,} Monthly Notices of the Royal Astronomical Society, 525, 2864, \dodoi{10.1093/mnras/stad2411}

\bibitem[{A. Oklopčić {et~al.}(2017)Oklopčić, Hopkins, Feldmann, Kereš, Faucher-Giguère, \& Murray}]{oklopcic_giant_2017}
Oklopčić, A., Hopkins, P.~F., Feldmann, R., {et~al.} 2017, \bibinfo{title}{Giant clumps in the {FIRE} simulations: a case study of a massive high-redshift galaxy,} Monthly Notices of the Royal Astronomical Society, 465, 952, \dodoi{10.1093/mnras/stw2754}

\bibitem[{ OpenAI(2024)OpenAI}]{openai_chatgpt_2024}
OpenAI. 2024, \bibinfo{title}{\{{ChatGPT}\}: {Chat}-based {Language} {Model},} \url{https://openai.com/chatgpt}

\bibitem[{I. Pasha \& T.~B. Miller(2023)Pasha \& Miller}]{pasha_pysersic_2023}
Pasha, I., \& Miller, T.~B. 2023, \bibinfo{title}{pysersic: {A} {Python} package for determining galaxy structural properties via {Bayesian} inference, accelerated with jax,} The Journal of Open Source Software, 8, 5703, \dodoi{10.21105/joss.05703}

\bibitem[{M. Puech {et~al.}(2009)Puech, Hammer, Flores, Neichel, \& Yang}]{puech_forming_2009}
Puech, M., Hammer, F., Flores, H., Neichel, B., \& Yang, Y. 2009, \bibinfo{title}{A forming disk at z {\textasciitilde} 0.6: collapse of a gaseous disk or major merger remnant?} Astronomy and Astrophysics, 493, 899, \dodoi{10.1051/0004-6361:200810521}

\bibitem[{D. Puskás {et~al.}(2025)Puskás, Tacchella, Simmonds, Hainline, D'Eugenio, Alberts, Arribas, Baker, Bunker, Carniani, Charlot, Duan, Eisenstein, Ji, Johnson, Jones, Maiolino, McClymont, Rieke, Rinaldi, Robertson, Übler, Williams, Willmer, Willott, \& Witstok}]{puskas_constraining_2025}
Puskás, D., Tacchella, S., Simmonds, C., {et~al.} 2025, \bibinfo{title}{Constraining the major merger history of z {\textasciitilde} 3-9 galaxies using {JADES}: dominant in situ star formation,} Monthly Notices of the Royal Astronomical Society, 540, 2146, \dodoi{10.1093/mnras/staf813}

\bibitem[{M.~J. Rieke {et~al.}(2023{\natexlab{a}})Rieke, Robertson, Tacchella, Hainline, Johnson, Hausen, Ji, Willmer, Eisenstein, Puskás, Alberts, Arribas, Baker, Baum, Bhatawdekar, Bonaventura, Boyett, Bunker, Cameron, Carniani, Charlot, Chevallard, Chen, Curti, Curtis-Lake, Danhaive, DeCoursey, Dressler, Egami, Endsley, Helton, Hviding, Kumari, Looser, Lyu, Maiolino, Maseda, Nelson, Rieke, Rix, Sandles, Saxena, Sharpe, Shivaei, Skarbinski, Smit, Stark, Stone, Suess, Sun, Topping, Übler, Villanueva, Wallace, Williams, Willott, Whitler, Witstok, \& Woodrum}]{rieke_jades_2023}
Rieke, M.~J., Robertson, B., Tacchella, S., {et~al.} 2023{\natexlab{a}}, \bibinfo{title}{{JADES} {Initial} {Data} {Release} for the {Hubble} {Ultra} {Deep} {Field}: {Revealing} the {Faint} {Infrared} {Sky} with {Deep} {JWST} {NIRCam} {Imaging},} The Astrophysical Journal Supplement Series, 269, 16, \dodoi{10.3847/1538-4365/acf44d}

\bibitem[{M.~J. Rieke {et~al.}(2023{\natexlab{b}})Rieke, Kelly, Misselt, Stansberry, Boyer, Beatty, Egami, Florian, Greene, Hainline, Leisenring, Roellig, Schlawin, Sun, Tinnin, Williams, Willmer, Wilson, Clark, Rohrbach, Brooks, Canipe, Correnti, DiFelice, Gennaro, Girard, Hartig, Hilbert, Koekemoer, Nikolov, Pirzkal, Rest, Robberto, Sunnquist, Telfer, Wu, Ferry, Lewis, Baum, Beichman, Doyon, Dressler, Eisenstein, Ferrarese, Hodapp, Horner, Jaffe, Johnstone, Krist, Martin, McCarthy, Meyer, Rieke, Trauger, \& Young}]{rieke_performance_2023}
Rieke, M.~J., Kelly, D.~M., Misselt, K., {et~al.} 2023{\natexlab{b}}, \bibinfo{title}{Performance of {NIRCam} on {JWST} in {Flight},} Publications of the Astronomical Society of the Pacific, 135, 028001, \dodoi{10.1088/1538-3873/acac53}

\bibitem[{ {Rieke, Marcia} {et~al.}(2023){Rieke, Marcia}, {Robertson, Brant}, {Tacchella, Sandro}, {Willmer, Christopher}, {Johnson, Ben}, {Carniani, Stefano}, {Bunker, Andy}, \& {Willott, Chris}}]{JADES-data}
{Rieke, Marcia}, {Robertson, Brant}, {Tacchella, Sandro}, {et~al.} 2023, \bibinfo{title}{Data from the {JWST} advanced deep extragalactic survey ({JADES}),} STScI/MAST, \dodoi{10.17909/8TDJ-8N28}

\bibitem[{P. Rinaldi {et~al.}(2025)Rinaldi, Rieke, Wu, Gilbert, Pacucci, Barchiesi, Alberts, Carniani, Bunker, Bhatawdekar, D'Eugenio, Ji, Johnson, Hainline, Kokorev, Kumari, Iani, Lyu, Maiolino, Parlanti, Robertson, Sun, Vignali, Williams, Willmer, \& Zhu}]{rinaldi_beyond_2025}
Rinaldi, P., Rieke, G.~H., Wu, Z., {et~al.} 2025, \bibinfo{title}{Beyond the {Dot}: an {LRD}-like nucleus at the {Heart} of an {IR}-{Bright} {Galaxy} and its implications for high-redshift {LRDs},} arXiv, \dodoi{10.48550/arXiv.2507.17738}

\bibitem[{A.~B. Romeo \& O. Agertz(2014)Romeo \& Agertz}]{romeo_larsons_2014}
Romeo, A.~B., \& Agertz, O. 2014, \bibinfo{title}{Larson's scaling laws, and the gravitational instability of clumpy discs at high redshift,} Monthly Notices of the Royal Astronomical Society, 442, 1230, \dodoi{10.1093/mnras/stu954}

\bibitem[{A.~B. Romeo {et~al.}(2010)Romeo, Burkert, \& Agertz}]{romeo_toomre-like_2010}
Romeo, A.~B., Burkert, A., \& Agertz, O. 2010, \bibinfo{title}{A {Toomre}-like stability criterion for the clumpy and turbulent interstellar medium,} Monthly Notices of the Royal Astronomical Society, 407, 1223, \dodoi{10.1111/j.1365-2966.2010.16975.x}

\bibitem[{J. Scholtz {et~al.}(2025)Scholtz, Carniani, Parlanti, D'Eugenio, Curtis-Lake, Jakobsen, Bunker, Cameron, Arribas, Baker, Charlot, Chevellard, Circosta, Curti, Duan, Eisenstein, Hainline, Ji, Johnson, Jones, Kumari, Maiolino, Maseda, Perna, Pérez-González, Rawle, Rieke, Rinaldi, Robertson, Saxena, Shivaei, Silcock, Sun, Rodríguez Del~Pino, Tacchella, Übler, Venturi, Williams, Willmer, Willott, \& Witstok}]{scholtz_jades_2025}
Scholtz, J., Carniani, S., Parlanti, E., {et~al.} 2025, \bibinfo{title}{{JADES} {Data} {Release} 4 -- {Paper} {II}: {Data} reduction, analysis and emission-line fluxes of the complete spectroscopic sample,} arXiv, \dodoi{10.48550/arXiv.2510.01034}

\bibitem[{T. Shibuya {et~al.}(2016)Shibuya, Ouchi, Kubo, \& Harikane}]{shibuya_morphologies_2016}
Shibuya, T., Ouchi, M., Kubo, M., \& Harikane, Y. 2016, \bibinfo{title}{Morphologies of {\textasciitilde}190,000 {Galaxies} at z = 0-10 {Revealed} with {HST} {Legacy} {Data}. {II}. {Evolution} of {Clumpy} {Galaxies},} The Astrophysical Journal, 821, 72, \dodoi{10.3847/0004-637X/821/2/72}

\bibitem[{V. Sok {et~al.}(2025)Sok, Muzzin, Tan, Asada, Bradač, Estrada-Carpenter, Iyer, Martis, Noirot, Sarrouh, Sawicki, Willott, Withers, Berek, \& Myers}]{sok_stellar_2025}
Sok, V., Muzzin, A., Tan, V. Y.~Y., {et~al.} 2025, \bibinfo{title}{The {Stellar} {Mass} and {Age} {Distributions} of {Star}-{Forming} {Clumps} at \$0.5 {\textless} z {\textless} 5\$ in {JWST} {CANUCS}: {Implications} for {Clump} {Formation} and {Destruction},} arXiv, \dodoi{10.48550/arXiv.2509.25363}

\bibitem[{S. Tacchella {et~al.}(2015)Tacchella, Carollo, Renzini, Förster~Schreiber, Lang, Wuyts, Cresci, Dekel, Genzel, Lilly, Mancini, Newman, Onodera, Shapley, Tacconi, Woo, \& Zamorani}]{tacchella_evidence_2015}
Tacchella, S., Carollo, C.~M., Renzini, A., {et~al.} 2015, \bibinfo{title}{Evidence for mature bulges and an inside-out quenching phase 3 billion years after the {Big} {Bang},} Science, 348, 314, \dodoi{10.1126/science.1261094}

\bibitem[{K.-i. Tadaki {et~al.}(2014)Tadaki, Kodama, Tanaka, Hayashi, Koyama, \& Shimakawa}]{tadaki_nature_2014}
Tadaki, K.-i., Kodama, T., Tanaka, I., {et~al.} 2014, \bibinfo{title}{The {Nature} of {Hα}-selected {Galaxies} at z {\textgreater} 2. {II}. {Clumpy} {Galaxies} and {Compact} {Star}-forming {Galaxies},} The Astrophysical Journal, 780, 77, \dodoi{10.1088/0004-637X/780/1/77}

\bibitem[{V. Tamburello {et~al.}(2017)Tamburello, Rahmati, Mayer, Cava, Dessauges-Zavadsky, \& Schaerer}]{tamburello_clumpy_2017}
Tamburello, V., Rahmati, A., Mayer, L., {et~al.} 2017, \bibinfo{title}{Clumpy galaxies seen in {H} α: inflated observed clump properties due to limited spatial resolution and sensitivity,} Monthly Notices of the Royal Astronomical Society, 468, 4792, \dodoi{10.1093/mnras/stx784}

\bibitem[{S. van~der Walt {et~al.}(2011)van~der Walt, Colbert, \& Varoquaux}]{van_der_walt_numpy_2011}
van~der Walt, S., Colbert, S.~C., \& Varoquaux, G. 2011, \bibinfo{title}{The {NumPy} {Array}: {A} {Structure} for {Efficient} {Numerical} {Computation},} Computing in Science and Engineering, 13, 22, \dodoi{10.1109/MCSE.2011.37}

\bibitem[{S. van~der Walt {et~al.}(2014)van~der Walt, Schönberger, Nunez-Iglesias, Boulogne, Warner, Yager, Gouillart, Yu, \& {scikit-image Contributors}}]{van_der_walt_scikit-image_2014}
van~der Walt, S., Schönberger, J.~L., Nunez-Iglesias, J., {et~al.} 2014, \bibinfo{title}{scikit-image: {Image} processing in {Python},} PeerJ, 2, e453, \dodoi{10.7717/peerj.453}

\bibitem[{A.~d.~l. Vega {et~al.}(2025)Vega, Mobasher, Manesh, Sharei, Chartab, \& Sattari}]{vega_fraction_2025}
Vega, A. d.~l., Mobasher, B., Manesh, F., {et~al.} 2025, \bibinfo{title}{The {Fraction} of {Clumpy} {Galaxies} in {JWST} {Surveys} over \$2{\textless}z{\textless}12\$,} arXiv, \dodoi{10.48550/arXiv.2508.14972}

\bibitem[{C.~C. Williams {et~al.}(2023)Williams, Tacchella, Maseda, Robertson, Johnson, Willott, Eisenstein, Willmer, Ji, Hainline, Helton, Alberts, Baum, Bhatawdekar, Boyett, Bunker, Carniani, Charlot, Chevallard, Curtis-Lake, de~Graaff, Egami, Franx, Kumari, Maiolino, Nelson, Rieke, Sandles, Shivaei, Simmonds, Smit, Suess, Sun, Übler, \& Witstok}]{williams_jems_2023}
Williams, C.~C., Tacchella, S., Maseda, M.~V., {et~al.} 2023, \bibinfo{title}{{JEMS}: {A} {Deep} {Medium}-band {Imaging} {Survey} in the {Hubble} {Ultra} {Deep} {Field} with {JWST} {NIRCam} and {NIRISS},} The Astrophysical Journal Supplement Series, 268, 64, \dodoi{10.3847/1538-4365/acf130}

\bibitem[{S. Wuyts {et~al.}(2012)Wuyts, Förster~Schreiber, Genzel, Guo, Barro, Bell, Dekel, Faber, Ferguson, Giavalisco, Grogin, Hathi, Huang, Kocevski, Koekemoer, Koo, Lotz, Lutz, McGrath, Newman, Rosario, Saintonge, Tacconi, Weiner, \& van~der Wel}]{wuyts_smoother_2012}
Wuyts, S., Förster~Schreiber, N.~M., Genzel, R., {et~al.} 2012, \bibinfo{title}{Smooth(er) {Stellar} {Mass} {Maps} in {CANDELS}: {Constraints} on the {Longevity} of {Clumps} in {High}-redshift {Star}-forming {Galaxies},} The Astrophysical Journal, 753, 114, \dodoi{10.1088/0004-637X/753/2/114}

\bibitem[{L. Yang {et~al.}(2025)Yang, Kartaltepe, Franco, Ding, Achenbach, Arango-Toro, Casey, Drakos, Faisst, Gillman, Gozaliasl, Huertas-Company, Jin, Liu, Magdis, Massey, Silverman, Tanaka, Yu, Akins, Allen, Ilbert, Koekemoer, McCracken, Paquereau, Rhodes, Robertson, Shuntov, \& Toft}]{yang_cosmos-web_2025}
Yang, L., Kartaltepe, J.~S., Franco, M., {et~al.} 2025, \bibinfo{title}{{COSMOS}-{Web}: {Unraveling} the {Evolution} of {Galaxy} {Size} and {Related} {Properties} at \$2,} arXiv, \dodoi{10.48550/ARXIV.2504.07185}

\bibitem[{A. Zanella {et~al.}(2015)Zanella, Daddi, Le~Floc'h, Bournaud, Gobat, Valentino, Strazzullo, Cibinel, Onodera, Perret, Renaud, \& Vignali}]{zanella_extremely_2015}
Zanella, A., Daddi, E., Le~Floc'h, E., {et~al.} 2015, \bibinfo{title}{An extremely young massive clump forming by gravitational collapse in a primordial galaxy,} Nature, 521, 54, \dodoi{10.1038/nature14409}

\bibitem[{A. Zanella {et~al.}(2019)Zanella, Le~Floc'h, Harrison, Daddi, Bernhard, Gobat, Strazzullo, Valentino, Cibinel, Sánchez~Almeida, Kohandel, Fensch, Behrendt, Burkert, Onodera, Bournaud, \& Scholtz}]{zanella_contribution_2019}
Zanella, A., Le~Floc'h, E., Harrison, C.~M., {et~al.} 2019, \bibinfo{title}{A contribution of star-forming clumps and accreting satellites to the mass assembly of z ∼ 2 galaxies,} Monthly Notices of the Royal Astronomical Society, 489, 2792, \dodoi{10.1093/mnras/stz2099}

\bibitem[{Y. Zhu {et~al.}(2025{\natexlab{a}})Zhu, Rieke, Ji, Simmonds, Sun, Sun, Alberts, Bhatawdekar, Bunker, Cargile, Carniani, de~Graaff, Hainline, Helton, Jones, Lyu, Rieke, Rinaldi, Robertson, Scholtz, Übler, Williams, \& Willmer}]{zhu_systematic_2025}
Zhu, Y., Rieke, M.~J., Ji, Z., {et~al.} 2025{\natexlab{a}}, \bibinfo{title}{A {Systematic} {Search} for {Galaxies} with {Extended} {Emission} {Lines} and {Potential} {Outflows} in {JADES} {Medium}-band {Images},} The Astrophysical Journal, 986, 162, \dodoi{10.3847/1538-4357/add2f4}

\bibitem[{Y. Zhu {et~al.}(2025{\natexlab{b}})Zhu, Bonaventura, Sun, Rieke, Alberts, Lyu, Shivaei, Morrison, Ji, Egami, Helton, Rieke, Rinaldi, Sun, \& Willmer}]{zhu_smiles_2025}
Zhu, Y., Bonaventura, N., Sun, Y., {et~al.} 2025{\natexlab{b}}, \bibinfo{title}{{SMILES} {Data} {Release} {II}: {Probing} {Galaxy} {Evolution} during {Cosmic} {Noon} and {Beyond} with {NIRSpec} {Medium}-{Resolution} {Spectra},} arXiv, \dodoi{10.48550/arXiv.2508.12599}

\bibitem[{Y. Zhu {et~al.}(2025{\natexlab{c}})Zhu, Alberts, Lyu, Morrison, Rieke, Sun, Helton, Ji, Bhatawdekar, Bonaventura, Bunker, Lin, Rieke, Rinaldi, Shivaei, Willmer, \& Zhang}]{zhu_higher_2025}
Zhu, Y., Alberts, S., Lyu, J., {et~al.} 2025{\natexlab{c}}, \bibinfo{title}{{SMILES}: {Potentially} {Higher} {Ionizing} {Photon} {Production} {Efficiency} in {Overdense} {Regions},} The Astrophysical Journal, 986, 18, \dodoi{10.3847/1538-4357/add263}

\bibitem[{H. Übler {et~al.}(2019)Übler, Genzel, Wisnioski, Förster~Schreiber, Shimizu, Price, Tacconi, Belli, Wilman, Fossati, Mendel, Davies, Beifiori, Bender, Brammer, Burkert, Chan, Davies, Fabricius, Galametz, Herrera-Camus, Lang, Lutz, Momcheva, Naab, Nelson, Saglia, Tadaki, van Dokkum, \& Wuyts}]{ubler_evolution_2019}
Übler, H., Genzel, R., Wisnioski, E., {et~al.} 2019, \bibinfo{title}{The {Evolution} and {Origin} of {Ionized} {Gas} {Velocity} {Dispersion} from z ∼ 2.6 to z ∼ 0.6 with {KMOS3D},} The Astrophysical Journal, 880, 48, \dodoi{10.3847/1538-4357/ab27cc}

\end{thebibliography}
\bibliographystyle{aasjournalv7}

\end{document}